\documentclass[prl,aps,groupedaddress]{revtex4}
\usepackage{amssymb,amsmath,amsfonts,epsfig,graphicx,latexsym,bm,color}

\begin{document}
\title{Orbital optical lattices with bosons}
\author{T. Kock$^{1}$, C. Hippler$^{1}$, A. Ewerbeck$^{1}$, and A. Hemmerich$^{1,2,3,}$\footnote{e-mail: hemmerich@physnet.uni-hamburg.de}}
\affiliation{$^{1}$Institut f\"{u}r Laser-Physik, Universit\"{a}t Hamburg, Luruper Chaussee 149, 22761 Hamburg, Germany}
\affiliation{$^{2}$Hamburg Centre for Ultrafast Imaging, Luruper Chaussee 149, 22761 Hamburg, Germany}
\affiliation{$^{3}$Wilczek Quantum Center, Zhejiang University of Technology, Hangzhou, China}

\date{\today}

\begin{abstract}
This article provides a synopsis of our recent experimental work exploring Bose-Einstein condensation in metastable higher Bloch bands of optical lattices. Bipartite lattice geometries have allowed us to implement appropriate band structures, which meet three basic requirements: the existence of metastable excited states sufficiently protected from collisional band relaxation, a mechanism to excite the atoms initially prepared in the lowest band with moderate entropy increase, and the possibility of cross-dimensional tunneling dynamics, necessary to establish coherence along all lattice axes. A variety of bands can be selectively populated and a subsequent thermalisation process leads to the formation of a condensate in the lowest energy state of the chosen band. As examples the 2nd, 4th and 7th bands in a bipartite square lattice are discussed. In the 2nd and 7th band, the band geometry can be tuned such that two inequivalent energetically degenerate energy minima arise at the $X_{\pm}$-points at the edge of the 1st Brillouin zone. In this case even a small interaction energy is sufficient to lock the phase between the two condensation points such that a complex-valued chiral superfluid order parameter can emerge, which breaks time reversal symmetry. In the 4th band a condensate can be formed in the $\Gamma$-point in the center of the 1st Brillouin zone, which can be used to explore topologically protected band touching points. The new techniques to access orbital degrees of freedom in higher bands greatly extend the class of many-body scenarios that can be explored with bosons in optical lattices.
\end{abstract}

\pacs{03.75.Lm, 03.75.Hh, 03.75.Nt} 

\maketitle

\section{Introduction}

The idea of trapping neutral atoms in periodic light shift potentials made by interfering laser beams has been conceived in the nineteen sixties as a possible means to localize the atoms below an optical wavelength and thus to overcome the notorious limitations of spectroscopy by Doppler broadening \cite{Let:68}. It was soon realized that even with the highest intensities conceivable with feasible laser technology the depth of the realizable potentials would be constrained to the mK range such that extraordinarily low temperatures would be required, unaccessible with state of the art cooling technology at that time. Hence, optical lattices remained science fiction until the concept of laser cooling was developed in the nineteen eighties, suddenly allowing sub-mK temperatures \cite{Jes:96, Ada:97, Met:99}. In the beginning nineteen nineties the first optical lattices with laser cooled atoms were prepared \cite{Ver:92, Jes:92, Hem1:93, Hem2:93, Gry:93} and immediately the regular arrangement of the atoms was made visible with the technique of Bragg scattering, well known from X-ray crystallography \cite{Wei:95, Bir:95}. A new vision was born, namely that such volatile systems of thin gases spatially ordered in a web of light could mimic crystalline solids. Unfortunately, temperatures were still too high to allow light-shift potentials sufficiently shallow to enable significant rates for tunneling between adjacent wells. This, however, appeared to be the crucial step to proceed from an array of independent microscopic atom traps to an ensemble with some collective character reminiscent of what is found in solids. Further cooling was soon achieved via evaporation \cite{Hes:86}, leading to the formation of Bose-Einstein condensates (BECs) \cite{And:95}, which could be readily loaded even into shallow light-shift potentials \cite{And:98}. Suddenly, the stage was prepared for the impressive success story of optical lattices that we have seen during the past two decades. Today, besides their well established relevance as a quantum laboratory for many-body lattice physics \cite{Lew:07}, also their originally envisioned objective as a tool for precision spectroscopy has become reality in their application in modern atomic clocks \cite{Der:11}.

An early milestone, pointing out the relevance of optical lattices for quantum many-body physics, has been the insight that optical lattices with bosons in their ground state represent a nearly ideal implementation of the bosonic Hubbard model \cite{Fis:89, Jak:98}. This has led to the demonstration of the superfluid-to-Mott-Insulator transition, when upon deepening the wells the system switches from the tunneling-dominated regime to the interaction-dominated regime \cite{Gre:02}. Despite this intriguing phenomenon of a phase transition triggered by quantum fluctuations, a notable limitation of the prospects of optical lattices for emulating more complex lattice models has soon become apparent. For bosonic atoms the ground states possible in optical lattices are necessarily positive definite, a very general property of bosonic ground state wave functions early discussed by Feynman \cite{Fey:72, Wu:09}. A possible way out of this dilemma is to amend the light shift potentials with artificial gauge fields \cite{Dal:11}, such that hopping processes are endowed with tailor-made phase factors or even more complex unitary transformations. Another obvious avenue, which is the topic of this article, is to utilize orbital degrees of freedom in higher Bloch bands. This strategy, geared by the paradigm of electronic matter, where orbital physics is ubiquitous (prominent examples are transition metal oxides \cite{Tok:00, Mae:04}), brings important new elements to the toolbox of optical lattices. Bosonic wave functions may be composed of high-order orbitals with complex nodal geometries with all the notorious extra freedom due to orientation, hybridization or energy degeneracy. The latter most importantly can come with the consequence that much lower energy scales than that of tunneling (for example weak collisional interactions) can become determinant for the structure of the minimal-energy state, a scenario known from quantum Hall physics. This may result in the emergence of unconventional chiral superfluid order in configuration space, which has a momentum space analogy in the global chiral symmetry of the superfluid order parameter proposed for the Cooper pairs of certain electronic superconductors \cite{Mae:94, Mae:04}. The purpose of this synopsis is to assemble the basic ideas, requirements and observations in our investigation of orbital physics in optical lattices with bosons. Experimental findings, distributed over a number of original articles, are combined with more recent, improved experimental data and extended theoretical interpretation in order to provide a convincing picture of our research results, accessible also for a non-specialist readership. So far, only a few experimental groups have engaged to explore metastable condensates in higher bands and many possibilities are still ahead. This brief review aims to stimulate further experimental groups to step in.   

A number of theory groups have pioneered our understanding of the physics of bosons condensed in higher bands. In particular, the physics of $p$-band bosons in a monopartite square lattice has been extensively studied in seminal work by Isacsson and Girvin in 2005 \cite{Isa:05} and by Liu and Wu in 2006 \cite{Liu:06}. In this work it has been pointed out that complex-valued, time reversal symmetry breaking superfluid order can emerge as a consequence of interactions between degenerate orbitals characterized by an orbital Hund's rule. A review of this early work is found in Ref. \cite{Wu:09}. More recently, this topic has received extensive interest in the condensed matter theory community leading to a wealth of proposals on orbital bosons and fermions in lattices with triangular, hexagonal and square geometries \cite{Cai:11, Li:11, Cai:12, Li:12, Mar:12, Sun:12, Heb:13, Liu:13, Pin:13, Li:14}.

\begin{figure}
\includegraphics[scale=0.5, angle=0, origin=c]{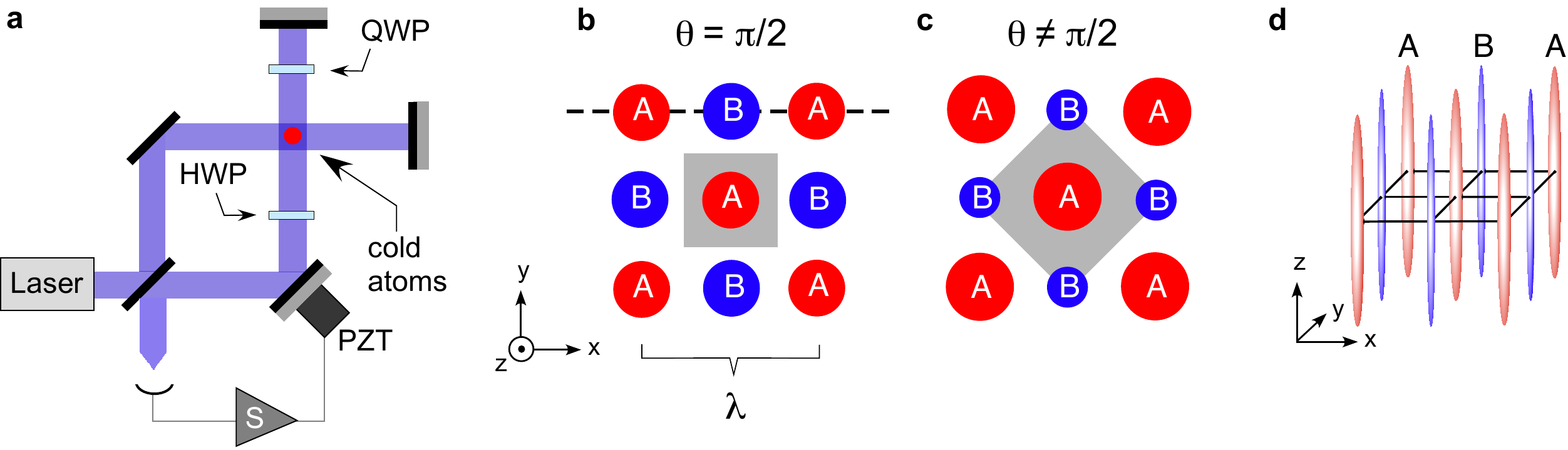}
\caption{(a) Sketch of the lattice set-up. Two optical standing waves with linear polarizations parallel to the $z$-axis are superimposed. HWP = half wave plate, QWP = quater wave plate, PZT= mirror mounted on piezo-electric transducer. S= servo control.  The lattice comprises two classes of wells denoted $\mathcal{A}$ and $\mathcal{B}$. Their relative depths can be tuned by adjustment of the phase angle $\theta$ in Eq.~\ref{Eq.1} via use of the PZT-mounted mirror in (a). In (b) the specific case of equal $\mathcal{A}$- and $\mathcal{B}$-wells is sketched. The general case is shown in (c). The grey areas denote the unit cells of the lattice. (d) If no additional lattice is applied along the $z$-direction, the atoms are weakly confined in this dimension, which gives rise to tubular lattice sites. 
\label{Fig.1}}
\end{figure}

An early experimental attempt to populate higher bands in optical lattices has been based upon the use of Raman excitation in a conventional square lattice with two additional laser beams oriented along two primitive vectors of the Bravais lattice \cite{Mue:07}. Although transfer of a fraction of the atoms into the second band was possible, the subsequent intra-band relaxation dynamics showed only little tendency to restore coherence. This experiment has taught us that a successful strategy to achieve condensates in higher bands relies on the interplay of a number of basic requirements, summarized in the following considerations. Long-lived condensates can, if at all, only be prepared in the lowest energy states of a band. These minimal energy states of higher bands typically have complex orbital geometries such that an efficient selective coherent transfer from the basically structureless condensate wave function in the ground state of the lowest band appears hopeless, since it would require the engineering of exceedingly complex excitation operators. Hence, the best possible choice is an excitation process, which populates a mixture of Bloch states energetically close to the minimal energy state of the target band. If the energy spread introduced in the excitation is kept sufficiently low, tunneling and elastic collisions may allow for a re-thermalisation of the sample with a significant condensate fraction in the minimal energy state and a thermal population of higher energy states within the band. A number of challenges come with this consideration. 1. The minimal energy state and the states nearby should be long-lived and in particular stable against two-body collisions. 2. An excitation mechanism is required, which can populate Bloch states sufficiently close to the minimal energy state of some band. 3. Tunneling should be equally possible in each direction within the lattice, in order to enable the formation of cross-dimensional coherence in a thermalisation process. As discussed in the following sections, a solution to these three requirements turns out possible in a bipartite square lattice with a rapidly tunable unit cell.

\begin{figure}
\includegraphics[scale=0.5, angle=0, origin=c]{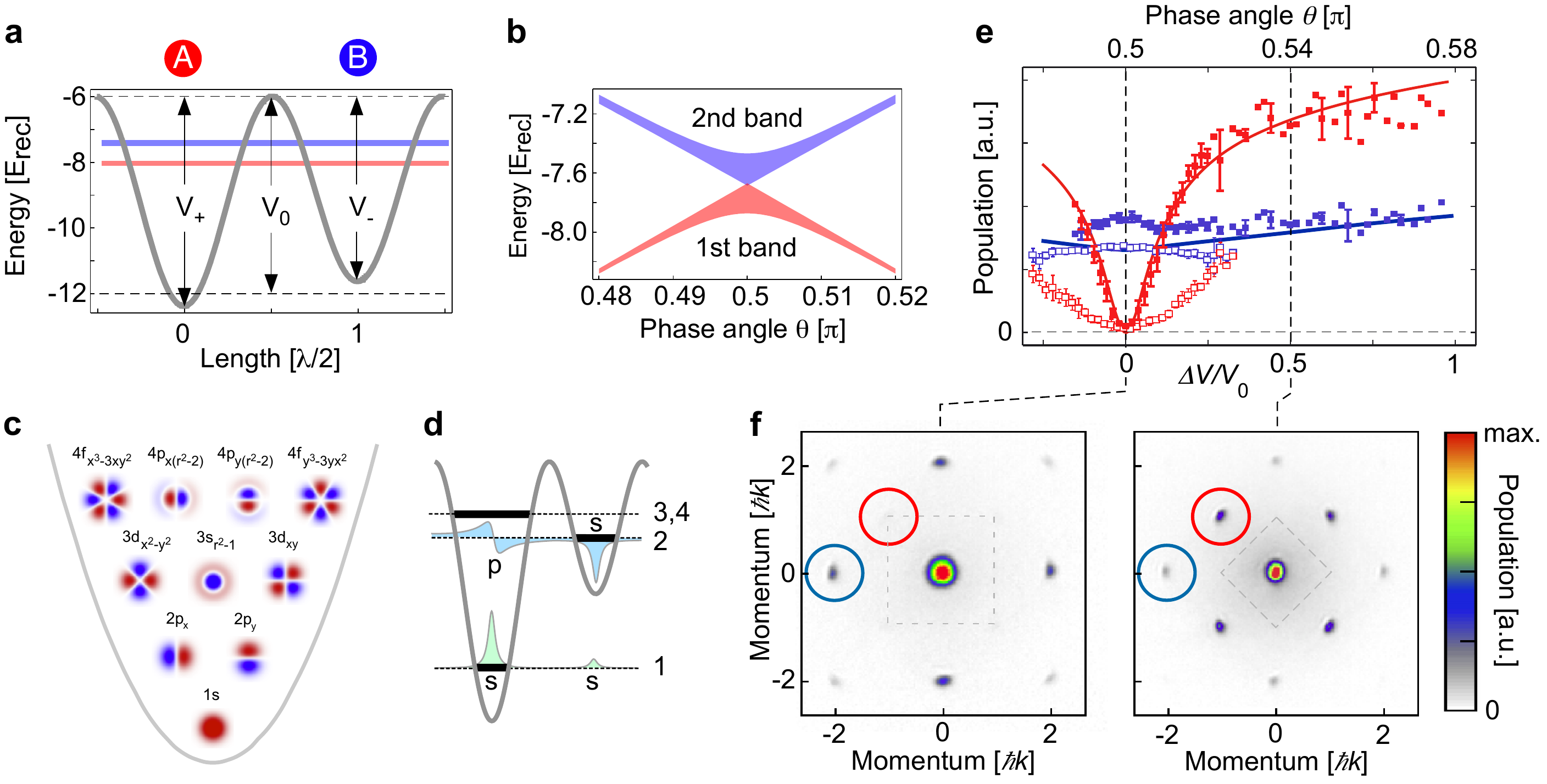}
\caption{\label{Fig.2} (a) The potential along the dashed horizontal trajectory in Fig.~\ref{Fig.1}(b) is plotted for $\theta = 0.51\, \pi$ and $V_{xy,0} = 6\,E_{\textrm{rec}}$ (thick grey line) with the first  and second bands represented, respectively, by the red and blue horizontal bars. (b) The first two bands are plotted versus $\theta$ for $V_{xy,0} = 6\,E_{\textrm{rec}}$. (c) Orbitals of the 2D harmonic oscillator. (d) If the $p_x$ and $p_y$-orbitals of the deep wells are tuned energetically close to the local $s$-orbitals in the shallow wells, the Bloch wave functions are composed of $s$-orbitals in both types of wells in the lowest band, and of $s$-orbitals and a superposition of $p_x$ and $p_y$-orbitals in the first excited band, respectively. (e) The red and blue squares show the relative number of atoms (normalized to the total particle number and plotted versus $\Delta V / V_{xy,0}$) associated with the Bragg peaks enclosed by red and blue circles in (f), respectively. The filled (open) squares are recorded for $V_{z,0}=0$ ($V_{z,0}= 22 \,E_{\textrm{rec}}$). The error bars indicate the statistical errors for five measurements. The solid lines are determined by a full band calculation (neglecting interaction) with no adjustable parameters. (f) Momentum spectra ($V_{xy,0} = 6 E_{\textrm{rec}}$) are shown with $\Delta V=0$ (left) and $\Delta V/V_{xy,0}=0.5$ (right) with the respective 1st BZs imprinted as dashed rectangles.}
\end{figure}

\section{Lattice potential}
Light-shift potentials with rapidly tunable lattice geometry lend themselves in various ways to control tunneling properties or to populate excited bands. Such potentials may be readily formed by superimposing lattices with different stationary geometries and rapidly tunable relative intensities \cite{Seb:06, Foe:07, And:07}. A particularly simple and versatile technique, which dates back to the pre-BEC era of optical lattices, relies upon interferometric phase control \cite{Hem:91, Hem:92, Hem1:93}. As we will see, this technique turns out tailor-made to realize the requirements for the implementation of orbital condensates. Using the interferometric lattice set-up illustrated in Fig.~\ref{Fig.1}(a), a two-dimensional (2D) optical potential is formed, which comprises two classes of wells ($\mathcal{A}$ and $\mathcal{B}$ in Fig.~\ref{Fig.1} (b,c)) arranged as the black and white fields of a chequerboard with an average well depth $V_{xy,0}$ (with regard to the $xy$-plane) and an adjustable relative potential energy offset $\Delta V$ between the two classes of sites. In the $xy$-plane the optical potential is given by
\begin{eqnarray}
\label{Eq.1}
V_{xy}(x,y) \,\equiv -\frac{V_{xy,0}}{4} \,  | \, \eta \, \left(e^{i k x}  + \epsilon_{x} \,e^{-i k x} \right) + \, e^{i \theta} \left(e^{i k y} + \epsilon_{y} \, e^{-i k y} \right) |^2 \,.
\end{eqnarray}
Here, $k \equiv 2 \pi/ \lambda$, and $\lambda = 1064$~nm is the wavelength of the lattice beams. Adjustment of $\theta$ with a precision exceeding $\pi/300$ (via servo-controlling the position of the PZT-mirror shown in Fig.~\ref{Fig.1}(a)) permits controlled tuning of $\Delta V \equiv V_{xy,0} \,\eta \,(1+\epsilon_{x}) (1+\epsilon_{y}) \cos(\theta)$ (cf. Fig.~\ref{Fig.1} (b,c)). If $\eta = \epsilon_{x} = \epsilon_{y}=1$, the lattice potential possesses perfect $C_4$ rotation symmetry. In various experiments reviewed here, due to unavoidable imperfections of the lattice set-up, one is constrained to fixed parameter values $\eta = 1.03$, and $\epsilon_{x} = 0.93$ and hence discrete $90^{\circ}$ rotation symmetry ($C_4$) is weakly broken. The optical set-up permits controlled adjustment of arbitrary values of $\epsilon_{y}$ within an interval $[0.9,1.1]$ including $\epsilon_{y}=1$. This is accomplished as follows: the optical standing wave along the $y$-axis is obtained by a retro-reflected laser beam. By means of a retardation plate (HWP in Fig.~\ref{Fig.1} (a)), the linear polarization of the incoming beam can be rotated away from the $z$-axis, such that the $z$-component of the polarization, which exclusively contributes to the lattice potential, is reduced. After retro-reflection the polarization is rotated to precisely match with the $z$-direction (QWP in Fig.~\ref{Fig.1} (a)). Adjustment of $\epsilon_{y}=1$ does not yield true $C_4$-symmetry of the lattice, however, as we will discuss later, it is sufficient to restore the $C_4$-symmetry close to the energy minima of certain bands. Motion of the atoms along the $z$-direction is either merely confined by a weak nearly harmonic potential (with 40 Hz vibrational frequency), which leads to a 2D lattice of elongated tubular sites (see Fig.~\ref{Fig.1} (d)), or by an additional lattice potential $V_z(z) \,\equiv -V_{\rm{z},0} \,  \cos^2(k z)$. If $V_{\rm{z},0}$ is chosen sufficiently large, the motion along the $z$-direction can be frozen out.

For any combination of the parameter values $\eta, \epsilon_{x} , \epsilon_{y}, \theta$ the lowest band possesses a single non-degenerate energy minimum at the $\Gamma$-point in the center of the first Brillouin zone (BZ). Bosons condensed in the lowest band populate the vicinity of the $\Gamma$-point. Hence, small distortions of the potential do not play a significant role for the physics of such bosons and the potential may be approximated (setting $\eta = \epsilon_{x} = \epsilon_{y}=1$) as
\begin{eqnarray}
\label{Eq.2}
V_{xy}(x,y) = -V_{xy,0} \left[ \cos^2 (k x) +  \cos^2 (k y) + 2 \cos(\theta)  \cos (k x) \cos (k y)\right] \,.
\end{eqnarray}
The effective well depths of the deep and shallow wells are then $V_{\pm} \equiv V_{xy,0}\,(1\pm\cos(\theta))^2$ and their difference $\Delta V \equiv V_{+} - V_{-} = 4\, V_{xy,0} \cos(\theta)$ (see Fig.~\ref{Fig.2}(a)). Tuning of $\theta$ significantly affects the lattice geometry. In the special case $\theta = \pi / 2$ (Fig.~\ref{Fig.1}(b)) the $\mathcal{A}$- and $\mathcal{B}$-wells become equal and a conventional monopartite square lattice geometry arises with primitive vectors $\frac{1}{2} \lambda \, \hat{x}$, $\frac{1}{2} \lambda\,\hat{y}$ (where $\hat{x}, \hat{y}$ denote the unit vectors in $x$- and $y$-direction, respectively). For $\theta \neq \pi / 2$ (cf. Fig.~\ref{Fig.1}(c)) the unit cell becomes bipartite and the lattice is spanned by the primitive vectors $\frac{1}{2} \lambda \, (\hat{x}+ \hat{y})$, $\frac{1}{2} \lambda \, (\hat{x}-\hat{y})$. Around $\theta=\pi/2$ the effective mean well depth $\bar V_{xy,0} \equiv (V_{+} + V_{-}) / 2 $ $= V_{xy,0} \left[1+\cos^2(\theta) \right]$ is only weakly dependent on $\theta$. For example, within the interval $0.46 < \theta / \pi < 0.54$ one has $ \cos^2(\theta) < 0.015$ and hence ${\bar V}_{xy,0} \approx V_{xy,0}$. The structure and the effective width of the lowest band notably depend on $\theta$. Equal $\mathcal{A}$- and $\mathcal{B}$-wells for $\theta = \pi/2$ facilitate tunneling as compared to values $\theta \neq \pi / 2$, where the broad lowest band of the $\theta = \pi / 2$-lattice splits into two more narrow bands as is shown in Fig.~\ref{Fig.2}(b).

The basic structure of the Bloch bands associated with the lattice potential in Eq.~\ref{Eq.1} may be understood within a tight-binding picture. Each lattice site may be approximated by an isotropic 2D harmonic oscillator with local orbitals (Wannier functions) exhibiting degeneracies and spatial distributions as illustrated in Fig.~\ref{Fig.2}(c). Each local orbital in the $\mathcal{A}$- and $\mathcal{B}$-wells gives rise to a band such that the band structure crucially depends on the setting of $\Delta V$. By tuning $\Delta V$ such that orbitals in different wells possess similar energy, additional band degeneracies can be adjusted. The bipartite lattice geometry typically gives rise to hybridized bands, i. e., the wave functions associated with the lowest bands will be typically composed of a local $s$-orbital in the shallow wells and higher-order orbitals in the deep wells superimposed by some of the sub-groups of degenerate orbitals shown in Fig.~\ref{Fig.2}(c). For example, if the $p_x$ and $p_y$-orbitals of the deeper wells are tuned energetically close to the local $s$-orbital in the shallow wells (as illustrated in Fig.~\ref{Fig.2}(d)), one may expect a chequerboard arrangement with $s$-orbitals in both types of wells in the lowest band, and $s$-orbitals and $c_x p_x + c_y p_y$-orbitals superimposed by $p_x$ and $p_y$-orbitals with complex coefficients $c_x$, $c_y$ in the first excited band, respectively.

As will be discussed in the following sections, the specific feature of the lattice potential of Eq.~\ref{Eq.1} that makes it particularly suitable for exploring orbital physics is the existence of two classes of wells, which can be tuned such that their role is interchanged, i.e. the deep wells may be rapidly tuned to become the shallow wells and vice versa. Using similar interferometric set-ups, analogue functionality can be engineered also in more complex lattice geometries, as for example in face-centered square lattices known to occur in high-$T_c$ cuprate superconductors or in hexagonal lattices known from graphene. Even lattice geometries with more than two classes of independently tunable sites should be possible.

\begin{figure}
\includegraphics[scale=0.3, angle=0, origin=c]{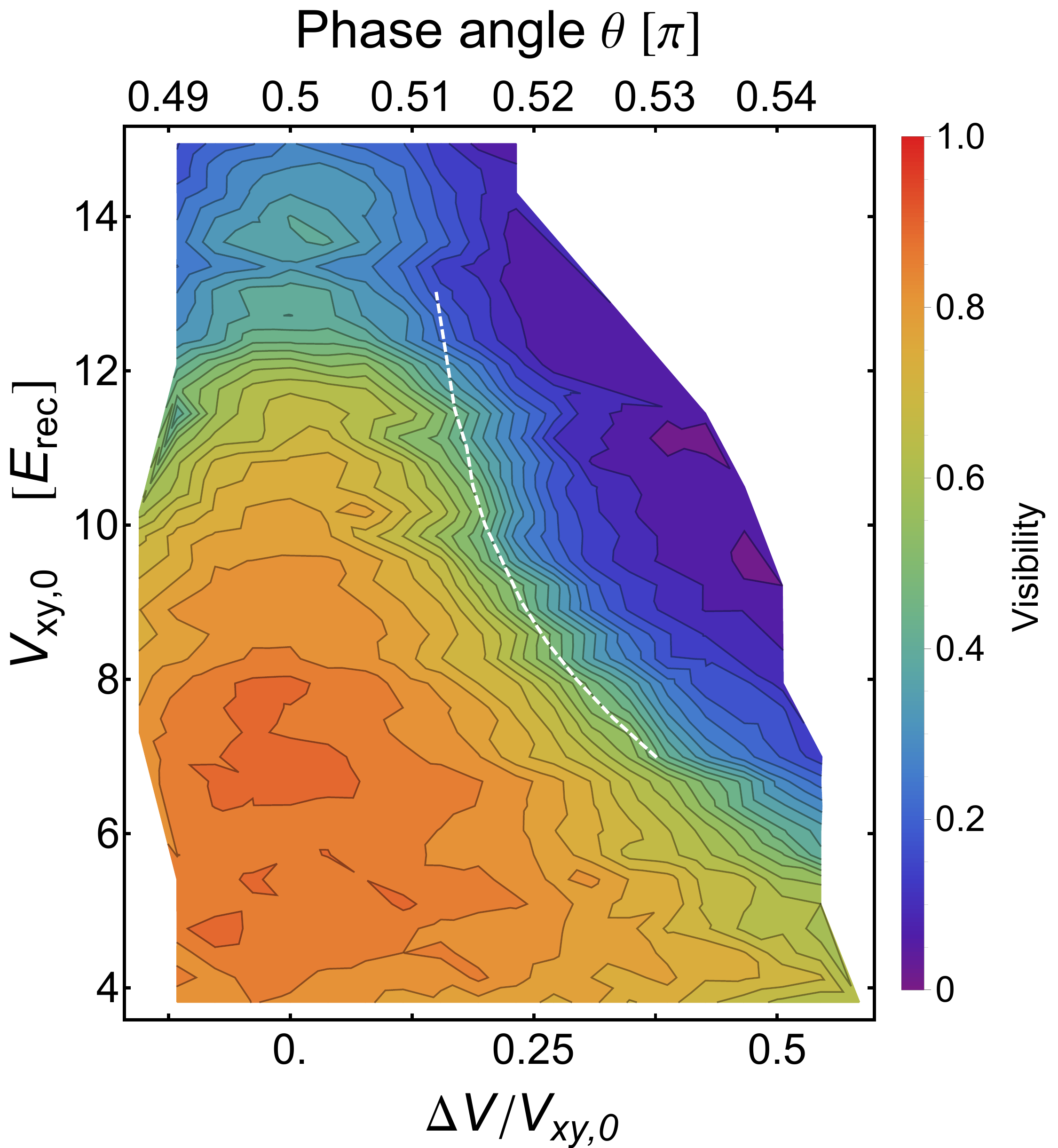}
\caption{The visibility of the $(-2,0) \hbar k$ Bragg peak, encircled blue in Fig.~\ref{Fig.2}(f), (parametrized by the color code shown on the right edge) is plotted as a function of the well depth parameter $V_{xy,0}$ (measured in units of the recoil energy $E_{\rm rec}$) and the potential energy off-set difference $\Delta V$ between shallow and deep wells in the bipartite lattice. The dashed line corresponds to the theoretical calculation of the points where the fraction of particles in the sublattice of shallow wells vanishes. Reproduced from Ref.~\cite{DiL:14}, \copyright\ Macmillan Publishers Ltd..}
\label{Fig.3}
\end{figure}

\section{Populating the lowest Bloch band}
The specific functionality of the lattice potential in Eq.~\ref{Eq.1} is readily tested by loading a BEC of $^{87}$Rb atoms into the lowest Bloch band. This is accomplished by forming a BEC with standard techniques \cite{Ket:99} and by means of a subsequent adiabatic increase of the overall potential depth $V_{xy,0}$ from zero to some desired value in typically 160~ms. For a well depth $V_{xy,0} = 6 \,E_{\textrm{rec}}$ ($E_{\textrm{rec}} = (\hbar k)^2/2m$ = recoil energy, $m$ = atomic mass) the dynamics in the $xy$-plane is dominated by tunneling such that the atoms form a coherent superfluid and one may record momentum spectra, which comprise pronounced Bragg maxima. Such spectra are readily produced by rapidly switching off all potentials and letting the atoms ballistically expand during 30 ms time of flight (TOF), after which an absorption image is recorded. The distribution of Bragg maxima reflects the shape of the underlying first BZ, which changes size and orientation as $\Delta V$ is detuned from zero. This is illustrated in Fig.~\ref{Fig.2}(e) and (f). In (f) two momentum spectra recorded for $\Delta V=0$ (left) and $\Delta V /V_{xy,0}=0.5$ (right) are shown with $V_{z,0} = 0$ in both cases. For $\Delta V=0$ (the special case of the monopartite square lattice of Fig.~\ref{Fig.1}(b)), the increased size of the 1st BZ gives rise to destructive interference, such that the $\pm (1,\pm 1)\,\hbar k$-Bragg peaks indicated by the red circles in (d) vanish. As $\Delta V$ is detuned from zero, a corresponding imbalance of the $\mathcal{A}$- and $\mathcal{B}$-populations yields a retrieval of the $\pm (1,\pm 1)\,\hbar k$-Bragg peaks. This is shown in Fig.~\ref{Fig.2}(e) for the case of approximately vanishing interaction energy per particle $U  \approx 0$ ($V_{\rm{z},0}=0$) by the filled red squares and for $U \approx  0.3 \,E_{\textrm{rec}}$ ($V_{\rm{z},0} = 22 \,E_{\textrm{rec}}$) by the open red squares, respectively. It is seen that the repulsive interaction significantly suppresses the formation of a population imbalance and corresponding $\pm (1,\pm 1)\,\hbar k$-Bragg peaks.

An early milestone experiment with optical lattices of bosons has been the demonstration of the superfluid-to-Mott-insulator transition \cite{Gre:02}. By increasing the overall well depth, the interaction energy per particle $U$ is increased and the hopping energy $J$ is decreased such that their ratio $U/J$ exceeds a critical value, where tunneling completely ceases and a shell structure of Mott insulators with different numbers of particles is formed in the external trap. This leads to a breakdown of coherence and hence the Bragg maxima in momentum spectra characteristic for the superfluid regime disappear. In the lattice discussed here, a superfluid-to-Mott-insulator transition can also be driven by tuning $\Delta V$, while the overall well depth $V_{xy,0}$ is kept constant. This resembles the possible control of coherence via a structural deformation of the unit cell in condensed matter, for example in high-$T_{c}$ cuprate systems \cite{DiL:14}. In Fig.~\ref{Fig.3} the visibility of the $(-2,0) \hbar k$ Bragg peak (for details see Ref.~\cite{DiL:14}) is plotted versus $V_{xy,0}$ and $\Delta V$. A pronounced plateau is seen in the lower left corner, where the system is coherent and superfluid. As $V_{xy,0}$ is increased along a vertical line with $\Delta V = 0$ a sharp decrease of visibility is seen, which corresponds to the superfluid-to-Mott-insulator transition studied in Ref.~\cite{Gre:02}. For constant $V_{xy,0}$, a steep decrease of the visibility is also induced as $\Delta V$ is tuned away from zero such that the unit cell acquires two wells of slightly different depth. The Mott insulator is formed when the populations in the shallow wells completely vanish, which is indicated by the white dashed line calculated in Ref.~\cite{DiL:14}.

\begin{figure}
\includegraphics[scale=0.5, angle=0, origin=c]{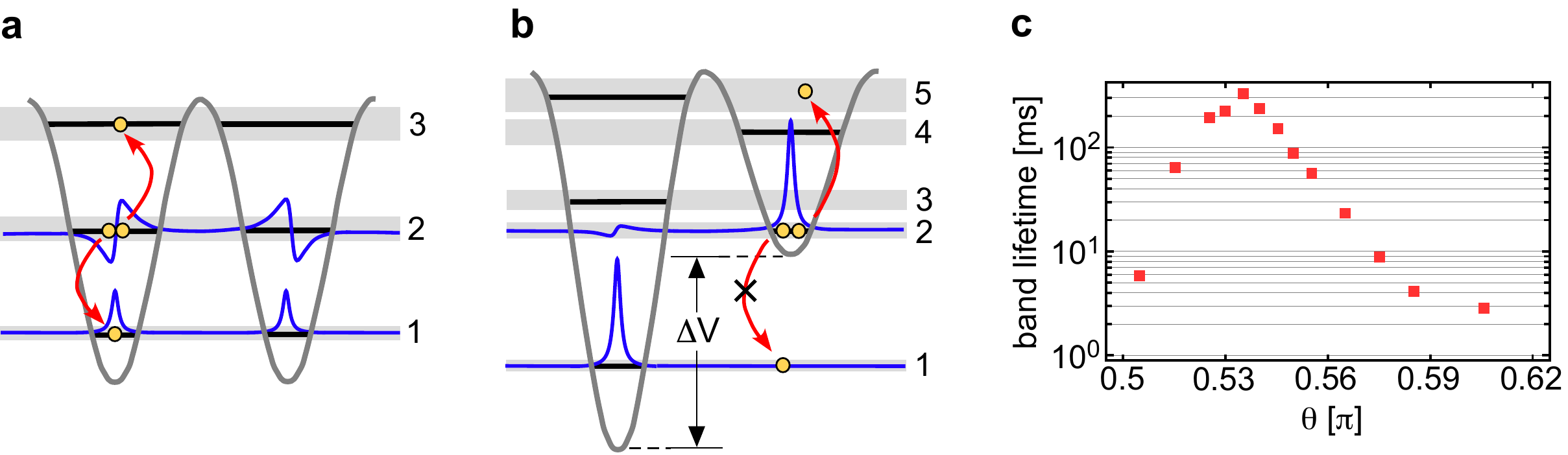}
\caption{(a) In a monopartite lattice, atoms colliding within the second band can be efficiently transferred to the first and third bands (red arrows). (b) In the second band of a bipartite lattice, inter-band collisions within the strongly populated local $s$-orbitals in the shallow wells are notably suppressed by a practically vanishing overlap with the ground state wave function (of the first band), which has a notable size only within the deep wells. The population in the local $p$-orbitals in the deep wells is so small that collision rates are low. In (a) and (b) the wave functions are sketched by the blue solid lines. The grey bars denote different bands associated with local states in the deep and shallow potential wells, indicated by solid lines within the grey bars. (c) Measured lifetime of the second band versus $\theta$ in Eq.~\ref{Eq.1} for $V_{xy,0} = 7\,E_{\textrm{rec}}$.}
\label{Fig.4}
\end{figure}

\section{Populating higher bands}

\subsection{Engineering long band lifetimes}
Higher bands are generally unstable with respect to binary collisions \cite{Isa:05}. For example, one of the collision partners may be transferred to a lower lying band while the other one is excited to a higher lying band. For this to be possible, energy momentum conservation requires the existence of target states at similar energy distances above and below the initially populated band. Hence, an appropriate design of the band gap structure allows one to control the collisional relaxation rates of higher bands to some extent \cite{Sto:08}. In the 2D lattice, discussed here, both collision partners may also be scattered to a lower lying band with the excess energy transferred to the dimension orthogonal to the lattice plane \cite{Bue:11, Pau:13}. Typical band lifetimes turn out to be on the ms time scale, which is often shorter than intra-band relaxation times. At sufficiently high densities, the discussion of band relaxation dynamics may be complicated by the possibility of dynamical instabilities \cite{Wu:01, Wu:03, Fal:04}. An important increase of the band lifetimes can arise if, in addition to energy and momentum conservation constraints, a spatial mismatch between the external wave functions of the initial states and the possible target states in other bands can be engineered. This is the case in bipartite lattices comprising deep and shallow potential wells. Such lattices give rise to hybrid bands composed of $s$-orbitals in the shallow wells and, for example, $p$-orbitals in the deep wells. For an appropriate adjustment of the relative well depth $\Delta V$, the local $p$-orbitals possess nearly vanishing stationary populations. Hence, notable collision rates only arise in the $s$-orbitals of the shallow wells. These have practically zero overlap with the ground state wave function in the lowest band, which is composed of local $s$-orbitals in the deep wells. This is illustrated in Fig.~\ref{Fig.4}, which compares the case of a monopartite lattice in (a) with that of a bipartite lattice in (b). In (c) a measurement of the band lifetime versus $\theta$ in Eq.~\ref{Eq.1} for $V_{xy,0} = 7\,E_{\textrm{rec}}$ is shown. A pronounced maximum is found around $\theta = 0.54\,\pi$, which corresponds to the case, where the population of the deep wells becomes minimal. A gain of band lifetime of nearly two orders of magnitude over the $\theta = 0.5\,\pi$ case can be obtained there. A theoretical study of band relaxation in a bipartite square lattice is found in Ref.~\cite{Pau:13}.

\begin{figure}
\includegraphics[scale=0.5, angle=0, origin=c]{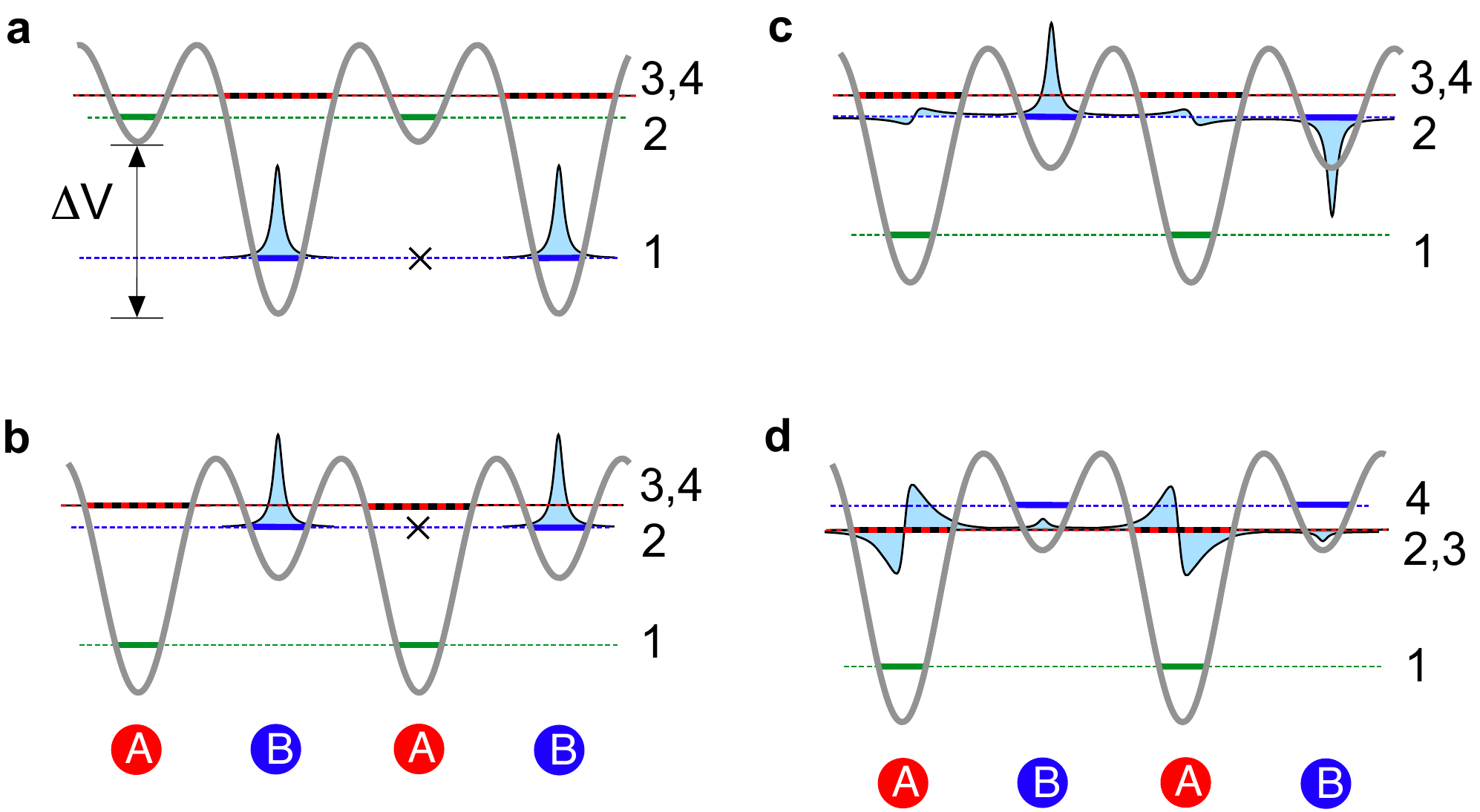}
\caption{The population of higher bands proceeds in four steps (a-d) detailed in the text. The thick grey solid lines show the optical potential along the dashed line in Fig.~\ref{Fig.1}(a). The thin colored dashed lines show the bands associated with the vibrational states in the deep and shallow wells indicated by thick solid lines of the same color. The light blue filled areas illustrate the local wave functions.}
\label{Fig.5}
\end{figure}

\subsection{Optimized excitation process}
Experiments generally start with a BEC adiabatically loaded into the lowest band of the lattice potential. An important requirement is to design the subsequent excitation process such that the states populated in the target band are energetically close to the minimal energy state of this band. An efficient direct excitation of the minimal energy state itself is not easily possible because this state typically possesses a complex spatial geometry with little overlap with the topologically trivial wave function of the ground state. In the following we discuss how bipartite lattices offer a straight forward method to efficiently populate higher bands with moderate heating. 

The experimental protocol, first explored in Ref.~\cite{Wir:11}, proceeds in four steps summarized in Fig.~\ref{Fig.5}. In step 1, the configuration illustrated in (a) is realized. After the preparation of the BEC in a magnetic trap, the lattice potential is slowly raised from zero to some final depth $V_{xy,0}$ in typically 160 ms such that all other dynamical time scales of the system are much shorter. The final values of $V_{xy,0}$ and $\Delta V < 0$ are chosen sufficiently large such that the atoms exclusively populate the lowest $s$-orbitals of the deep wells, and the relative phases between the local wave functions in different wells are nearly indeterminate due to interaction-driven squeezing of the particle number fluctuations. This lack of phase coherence is indicated in Fig.~\ref{Fig.5}(a) by the black cross between adjacent deep wells. In step 2, $\Delta V$ is tuned from the negative to the positive side, such that the deep wells become the shallow wells and vice versa. This leads to the configuration shown in (b). The time scale for this step is adjusted to be short with respect to tunneling (on the order of 1 ms) but long with respect to the on-site vibration period (on the order of 10 $\mu$s). This ensures that during this process the local wave function is not affected by tunneling, and that it adiabatically adjusts to the changing depth of the local potential well. Different target values of $\Delta V$ yield population of different bands. The precise value of $\Delta V$ is adjusted such that, for the lowest energy state in the target band, the population of the deep wells is minimized. This choice has two important consequences. It maximizes the band lifetime, as has been discussed in the previous section, and it gives rise to a state that is energetically close to the minimal energy state, such that a subsequent thermalisation process can lead to Bose-Einstein condensation in the band minima with a significant condensate fraction. For example, if the target band is the second band, the minimal energy wave function is comprised of local $s$-orbitals with relative local phases of $\pi$ in adjacent shallow wells, connected via $p$-orbitals in the deep wells, which are minimally populated if $\Delta V$ is chosen appropriately (see Fig.~\ref{Fig.5}(c)). Hence, the state produced after step 2 in (b) differs from the minimal energy state indicated in (c) only by the indeterminate phase relations between the $s$-orbitals and the missing populations in the $p$-orbitals. Note that the preparation of indeterminate phase relations in step 2 is energetically more favorable than that of wrongly fixed phase relations. The third step is to give the atoms time to tunnel and collide within the lattice, thus re-establishing thermal equilibrium. If the initial state prepared after step 2, has sufficiently low energy, this leads to a significant condensate fraction occupying the minimal energy state. In a final step, after the thermalisation process is completed, $\Delta V$ may be adiabatically tuned in order to prepare condensate wave functions with significant populations in the deep wells (the $p$-orbitals in (d)).

\begin{figure}
\includegraphics[scale=0.4, angle=0, origin=c]{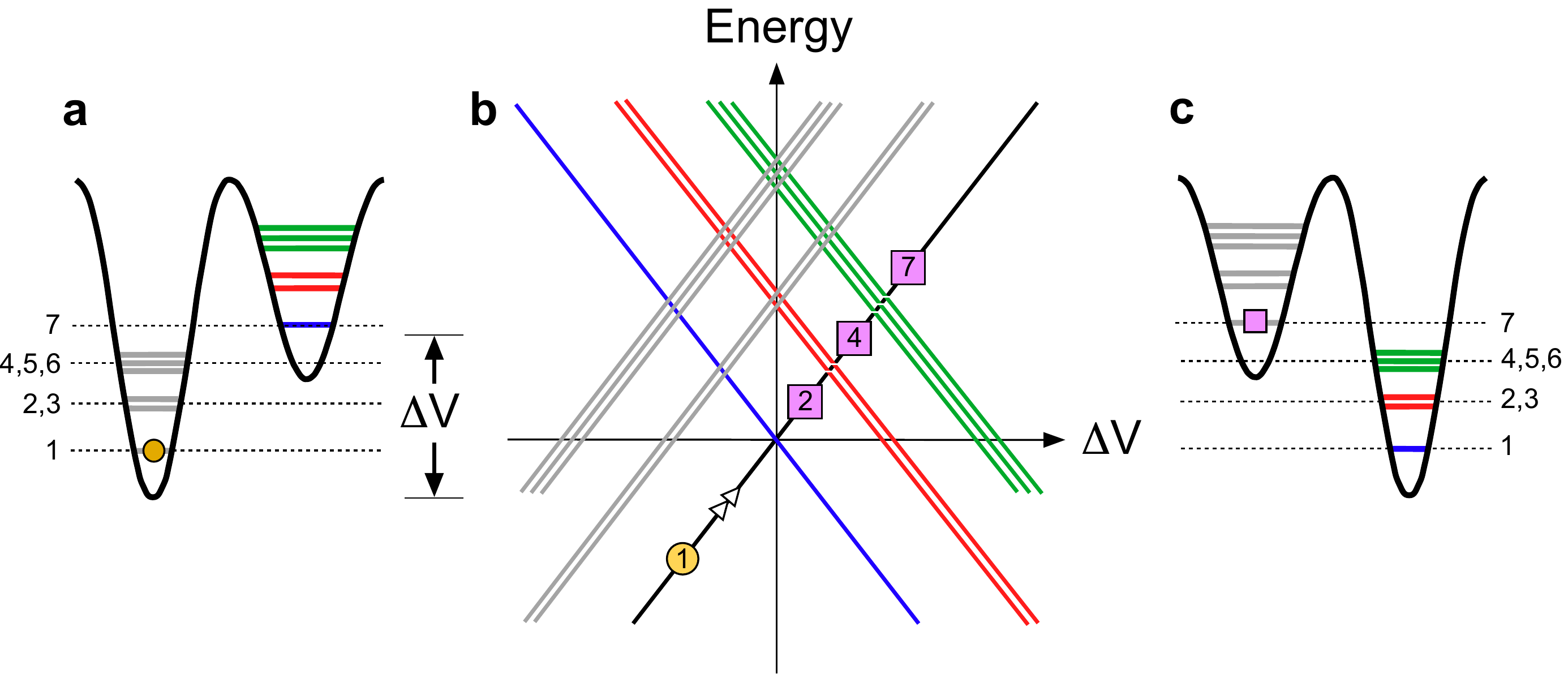}
\caption{The schematic evolution of the Bloch bands in a 2D bipartite lattice is shown in (b) as $\Delta V$ is tuned from the initial state sketched in (a) towards the final state sketched in (c). The orange disks in (a) and (b) indicate the initially populated band. The violet squares in (b) and (c) indicate possible bands populated at the end of the $\Delta V$ scan.}
\label{Fig.6}
\end{figure}

Not every Bloch band can be populated with the swapping technique sketched in Fig.~\ref{Fig.5}, as is discussed in Fig.~\ref{Fig.6}. The figure illustrates in (b) the evolution of the bands as the lattice is tuned from the initial state sketched in (a) to the final state sketched in (c). A 2D lattice is assumed here such that each potential well hosts local vibrational states indicated by a main quantum number $n \in \{1,2,...\}$ with approximate degeneracies $g_n = n$, each of which gives rise to a Bloch band. The bands with different $n$ are energetically separated by a significant amount, while those with the same $n$ are very closely spaced. It is then seen in (b) that, as $\Delta V$ is tuned from the negative to the positive side, the bands arising from the initially deep wells increase while those resulting from the initially shallow wells decrease in energy, which successively leads to weakly avoided band crossings. The $\Delta V$ scan is sufficiently fast that populations pass across these avoided crossings. As a consequence, the population of the initially lowest band (orange disk in (b)) can be transferred into the 2nd, the 4th, the 7th or some higher band (violet squares in (b)). The 3rd, 5th, or 6th bands cannot be selectively addressed because the gaps separating them from adjacent bands are too small. If the target value of $\Delta V$ is adjusted such that the lowest state in the shallow wells is tuned midway between two well separated groups of states with different values of $n$ in the deep wells, the resulting band possesses wave functions with little population in the deep wells. Hence this essential requirement for long band lifetimes and efficient condensation, discussed in the previous paragraph, can in fact be realized.

\begin{figure}
\includegraphics[scale=0.4, angle=0, origin=c]{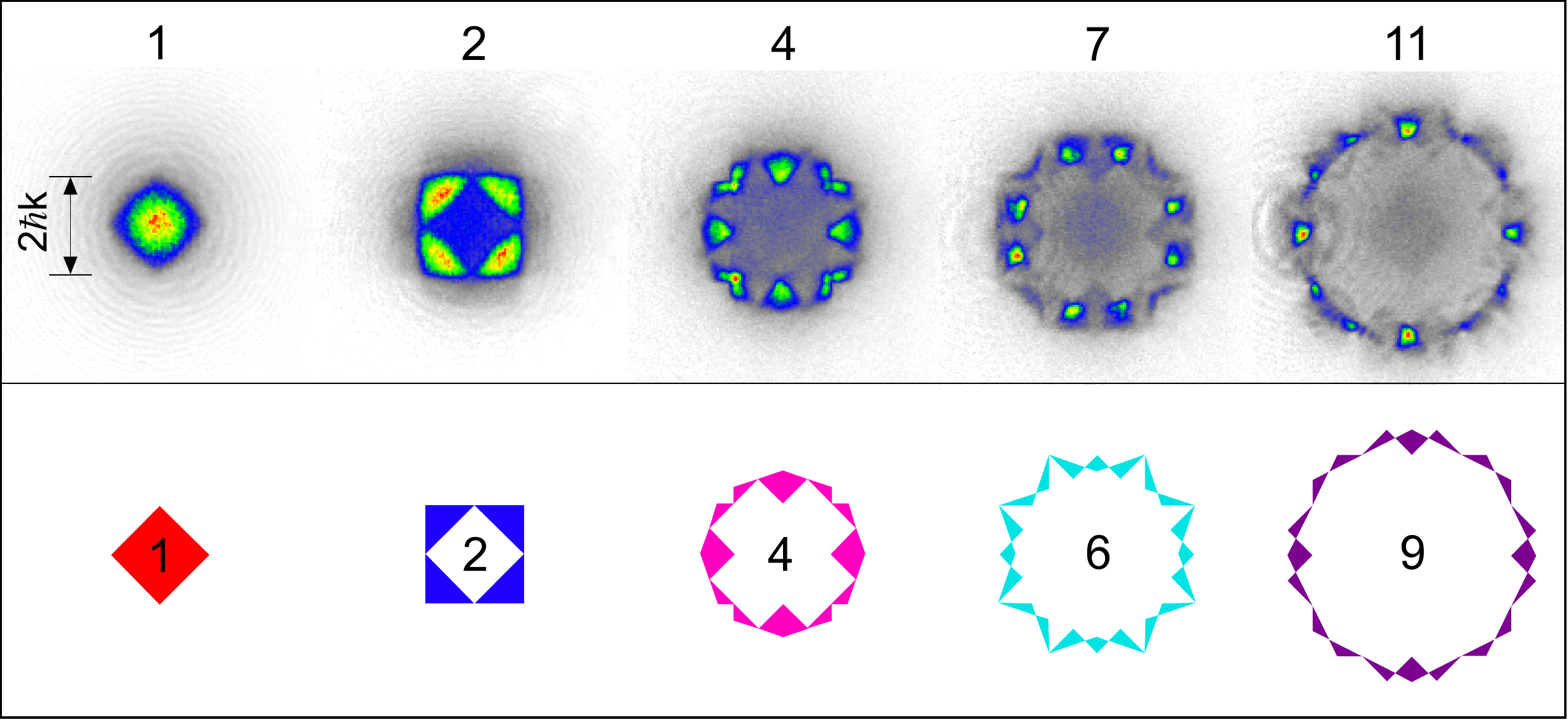}
\caption{The upper row shows (from left to right) experimentally observed momentum distributions without previous populating of higher bands and after applying the band swapping method of Fig.~\ref{Fig.5} optimized for population of the 2nd, the 4th, the 7th or the 11th band, respectively. The lower row depicts the theoretical shape of the most populated BZ indexed by the central number.}
\label{Fig.7}
\end{figure}

\begin{figure}
\includegraphics[scale=0.4, angle=0, origin=c]{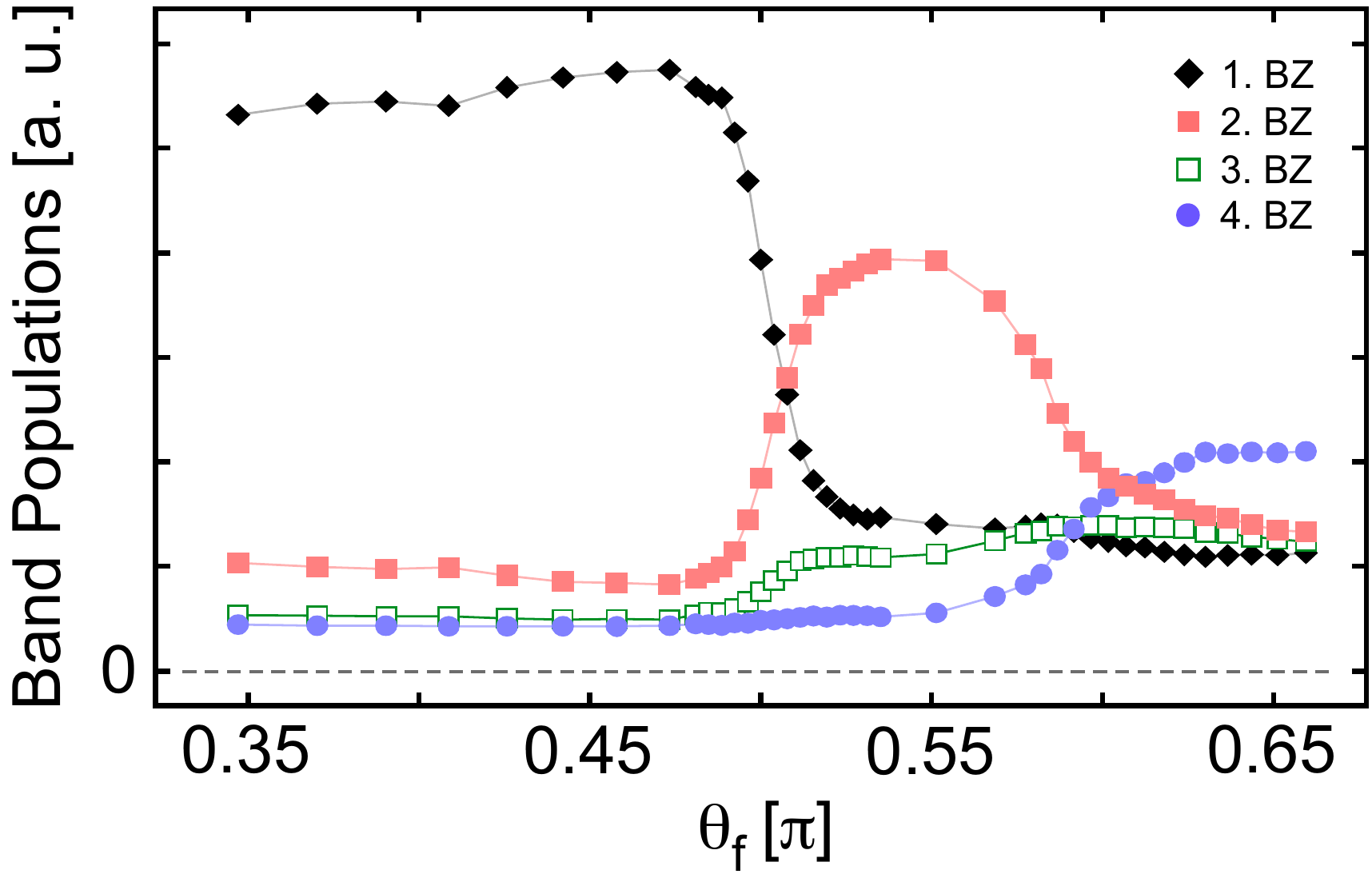}
\caption{Band populations observed for $V_{xy,0} = 7\,E_{\textrm{rec}}$ after applying the band swapping technique of Fig.~\ref{Fig.6}. The lattice is switched between an initial value $\theta_i = 0.35\,\pi$ and a variable final value $\theta_f$.}
\label{Fig.8}
\end{figure}

In order to probe the population of higher bands we employ a widely used band mapping technique. The lattice potential is adiabatically decreased within 2 ms such that the population of the n-th band is mapped onto the n-th BZ. A precondition is that as the lattice potential is ramped down, the bands maintain their energetic order, i.e. no band crossings occur. For the lowest bands this is usually the case. Via successive ballistic expansion during 30 ms, a density distribution of the atoms is formed and recorded by absorption imaging, which depicts momentum space. In Fig.~\ref{Fig.7} we present typical results of the band swapping technique used to populate the 2nd, the 4th, the 7th or the 11th band. The upper row shows experimental band mapping pictures of momentum space with the theoretical shape of the most populated BZs shown in the row below. While the 2nd and the 4th bands are in fact mapped onto the 2nd and 4th BZ, the 7th band is predominantly mapped onto the 6th BZ, because the 6th and 7th bands cross as the lattice depth is tuned to zero. Similarly, the 11th band is  predominantly mapped onto the 9th BZ. As suggested by Fig.~\ref{Fig.6}, significant selective population of the 3rd, 5th or 6th bands was not possible for any final setting of $\Delta V$. Fig.~\ref{Fig.8} shows a measurement of the band populations with $V_{xy,0} = 7\,E_{\textrm{rec}}$, an initial value $\theta_i = 0.35\, \pi$ corresponding to $\Delta V = -12.7\,E_{\textrm{rec}}$, and a variable final value $\theta_f$. The symbols show the populations of the first four bands obtained from counting the atoms in the respective BZs using band mapping spectra as shown in Fig.~\ref{Fig.7}. Depending on the value of $\theta_f$ the 2nd (around $\theta_f = 0.55 \,\pi$) and the 4th (around $\theta_f = 0.65 \,\pi$) bands can be predominantly populated, while this is not the case for the 3rd band in accordance with the expectations in the context of Fig.~\ref{Fig.6}.

\begin{figure}
\includegraphics[scale=0.6, angle=0, origin=c]{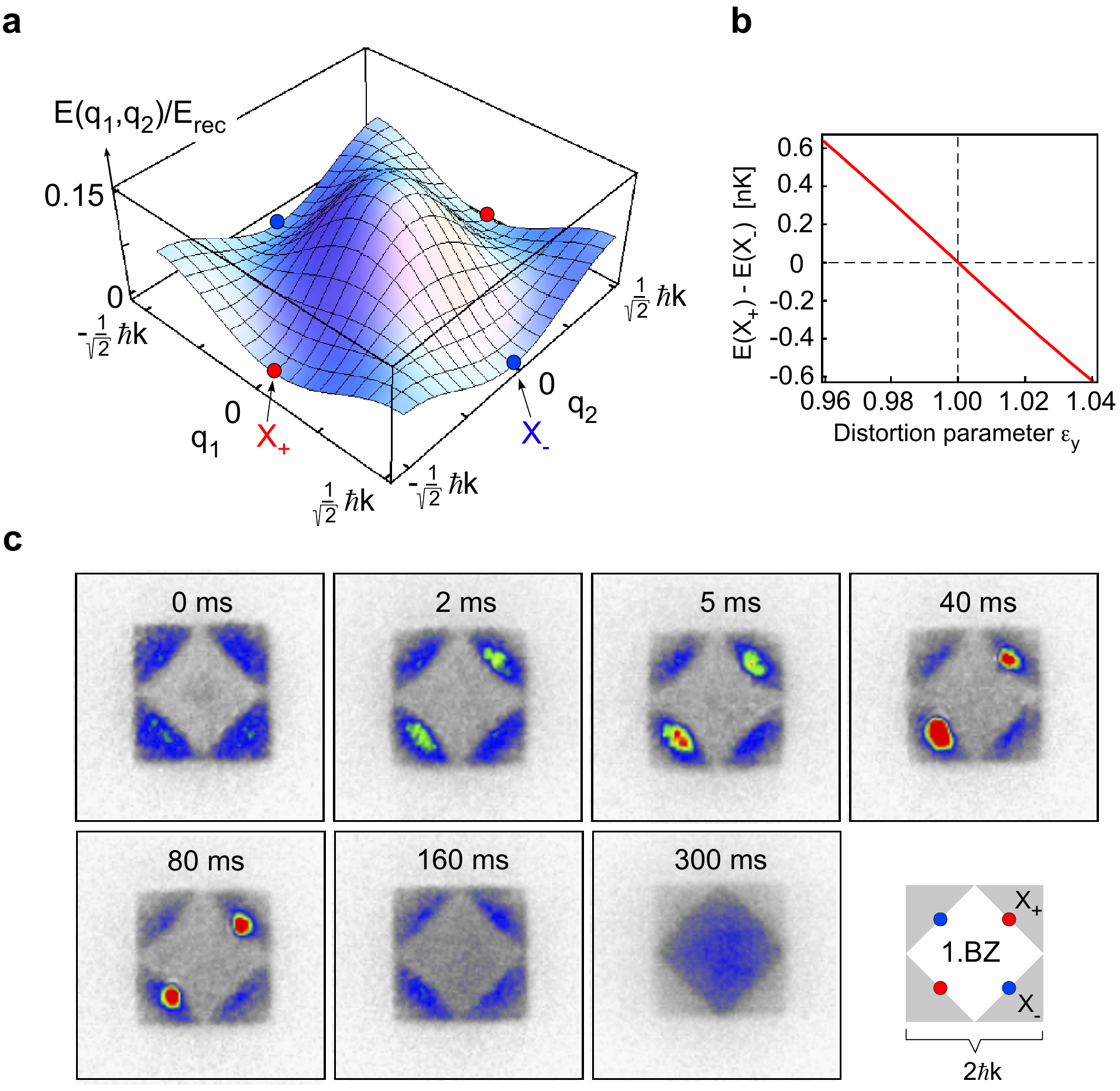}
\caption{(a) Energy surface $E(q_1,q_2)$ of the second band plotted across the 1st BZ. (b) The energy difference $E(X_{+}) - E(X_{-})$ between the condensation points $X_+$ and $X_-$, calculated from the potential in Eq.~\ref{Eq.1} with $V_{xy,0} = 7\,E_{\textrm{rec}}$ and $\theta = 0.57\,\pi$, is plotted in units of nanokelvin versus the distortion parameter $\epsilon_{y}$. (c) The formation of a condensate in the $X_+$-point is observed in a series of band mapping spectra recorded after variable thermalisation times indicated on the upper edge of the plots. Due to a small lattice distortion the energy of the $X_+$-point was lowered with respect to that of the $X_-$-point by 0.5 nanokelvin.}
\label{Fig.9}
\end{figure}

\begin{figure}
\includegraphics[scale=0.5, angle=0, origin=c]{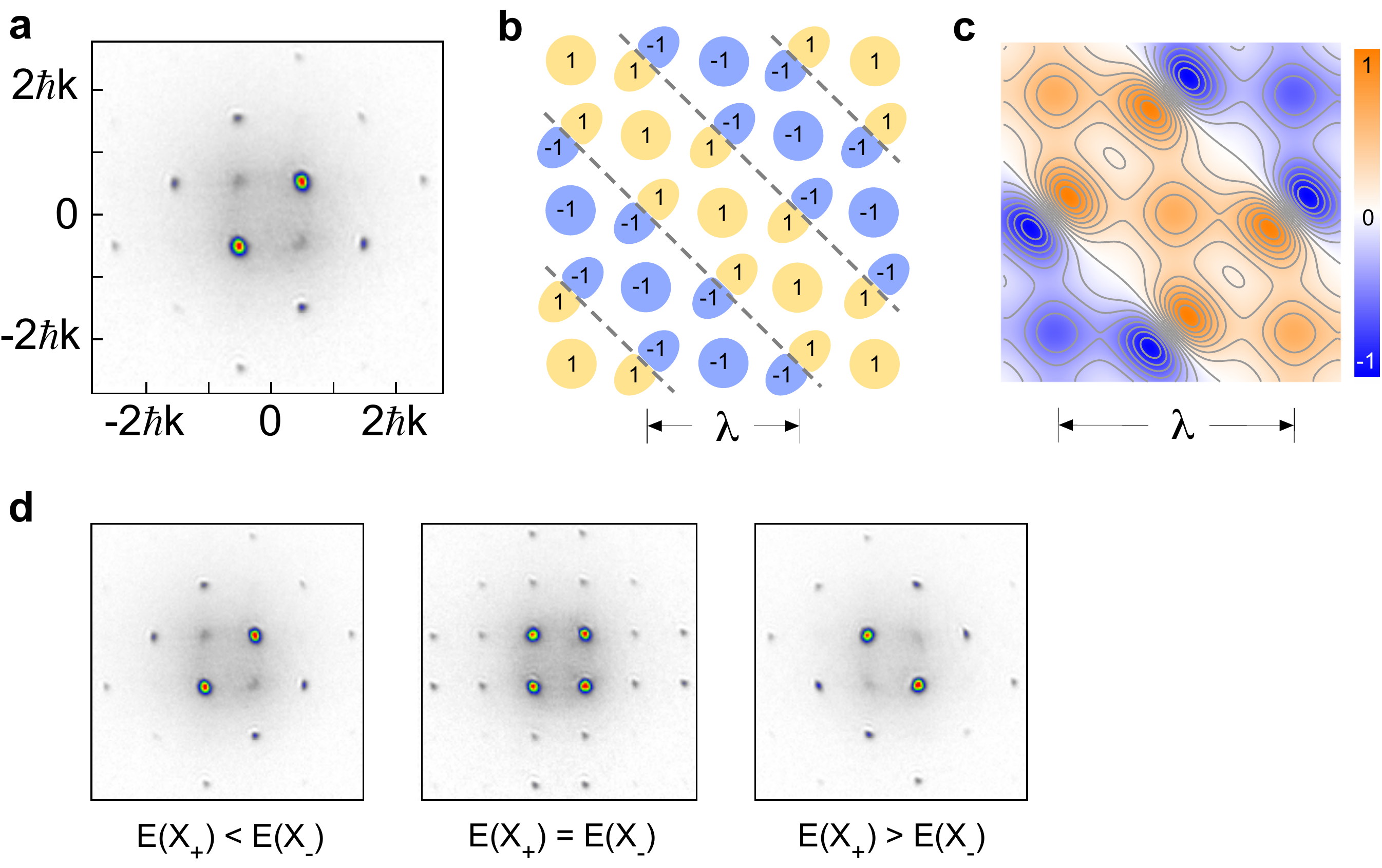}
\caption{(a) Momentum spectrum after holding the atoms in the lattice for 80 ms with $E(X_{+}) < E(X_{-})$. (b) Schematic of the spatial wave function corresponding to the spectrum in (a). (c) Bloch function associated with $X_{+}$, calculated for $V_{xy,0} = 7\,E_{\textrm{rec}}$ and $\theta = 0.75\,\pi$. (d) Momentum spectra for different settings of the condensation point energies $E(X_{\pm})$.}
\label{Fig.10}
\end{figure}

\section{Unconventional superfluid order in the second band}
In this section we focus on the physics associated with the 2nd Bloch band. In Fig.~\ref{Fig.9}(a) this band is plotted across the 1st BZ showing the two inequivalent energy minima at the edge of the 1st BZ (red and blue disks), which are denoted $X_{+}$ and $X_{-}$. Since the 2nd band hosts anisotropic $p$-orbitals in the deep wells, small distortions of the lattice, breaking its four-fold rotation symmetry ($C_4$), play an important role for the physics in this band. In particular, deviations from the $C_4$-symmetry can lead to a small energy separation of the two band minima, such that the calculation of the band requires to account for the full potential given in Eq.~\ref{Eq.1}. The energy difference between the condensation points $E(X_{+}) - E(X_{-})$ can be controlled on the sub-nanokelvin scale via the distortion parameter $\epsilon_{y}$ as shown in Fig.~\ref{Fig.9}(b). If bosonic atoms, prepared in the 2nd band at sufficiently low mean energy, are given time to thermally equilibrate, a fraction of them is expected to condense in the state of lowest energy, which is either of the $X_{\pm}$-points. That this is in fact the case is seen in the band mapping spectra shown in Fig.~\ref{Fig.9}(c). The atoms were excited to the 2nd band according to the procedure sketched in Fig.~\ref{Fig.5}(a) and (b) with $V_{xy,0} = 7\,E_{\textrm{rec}}$ and the final value of $\theta$ was set to $\theta_f = 0.54 \,\pi$. Subsequently, band mapping was applied after a variable holding time specified on the upper edge of the plots shown. The figure shows that initially the atoms occupy a larger region in the 2nd BZ. Through tunneling and collision dynamics a fraction of the atoms migrates to the $X_{+}$-point and a notable condensate fraction builds up within several tens of ms. Band relaxation due to collisions, which acts to refill the 1st BZ and hence the lowest band, does only become significant after about two hundred ms.

If the adiabatic decrease of the lattice potential used in the band mapping procedure is replaced by rapid switching, a conventional momentum spectrum is obtained. In case of a coherent atomic sample $|\psi_+\rangle$ in $X_+$ described by a wave function $\psi_{+}(r) \equiv \langle r |\psi_+\rangle$, the associated momentum spectrum is given by $|{\tilde \psi}_+(p)|^2$ with ${\tilde \psi}_+(p)$ denoting the Fourier transform of $\psi_+(r)$, which is composed of narrow Bragg resonances quantifying the harmonic components of $\psi_+(r)$. An example corresponding to the 80 ms band mapping spectrum in Fig.~\ref{Fig.9}(c) is shown in Fig.~\ref{Fig.10}(a). The absence of a Bragg peak at zero-momentum and hence a ($p=0$)-component in $|{\tilde \psi}_+(p)|^2$ shows that the associated spatial distribution $|{\psi}_+(r)|^2$ has standing-wave character. Obviously, a finite-momentum superfluid order has been formed. One may infer the structure of ${\psi}_+(r)$ by composing local $s$-orbitals in the shallow wells and local $p$-orbitals in the deep wells, arranging them such that the local phases at the tunneling bonds are equal in order to minimize kinetic energy. This is illustrated in Fig.~\ref{Fig.10}(b). A calculation of the Bloch function, shown in Fig.~\ref{Fig.10}(c), associated with the condensation point $X_+$, confirms this result. Similarly as for the ground state wave function in the lowest band, $\psi_{+}(r)$ can be chosen real, however in contrast to the ground state it exhibits nodal lines (dashed grey lines in Fig.~\ref{Fig.10}(b)).  

\begin{figure}
\includegraphics[scale=0.6, angle=0, origin=c]{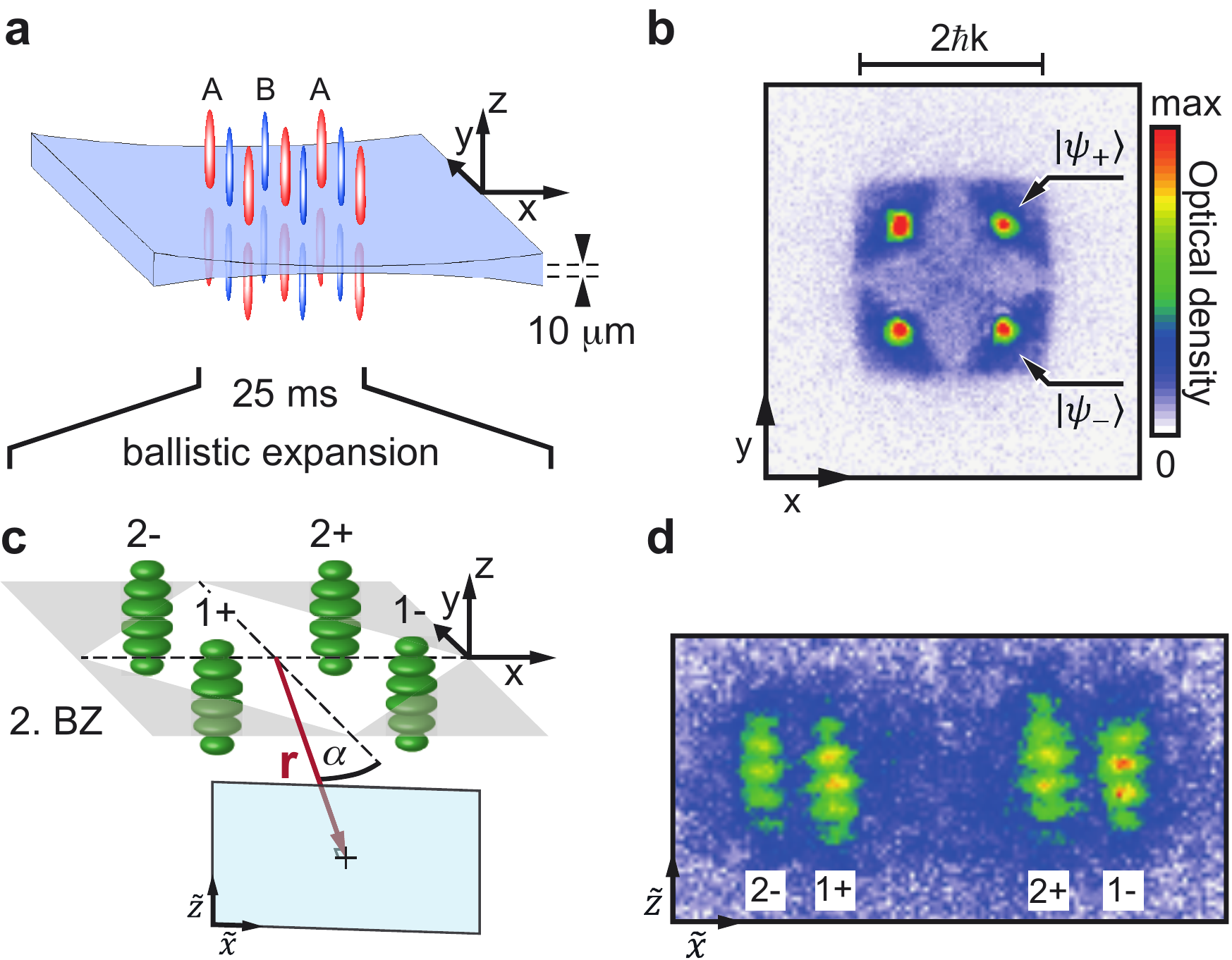}
\caption{(a) The tubular minima of the lattice potential are split into upper and lower sections with a cylindrically focused laser beam (light blue). (b) Band mapping image showing equally populated condensates at the minima of the second band $X_{\pm}$. (c) Sketch of the atomic spatial distribution after a ballistic expansion of $25\,$ms. The grey area corresponds to the second Brillouin zone. Each of the four Bragg maxima $B_{i, \sigma}$ (indexed $i \in \{1,2\}$, $\sigma \in \{-,+\}$), corresponding to the $X_\pm$-points in Fig.~\ref{Fig.9}(a), carries an interference grating aligned with the $z$-axis. (d) Absorption image along the line of sight indicated by the vector $\mathbf{r}$ in (c), which lies in the $xy$-plane and encloses an angle of $\alpha=13^\circ$ with the $y$-axis. Reproduced from Ref.~\cite{Koc:15}, \copyright\ American Physical Society.}
\label{Fig.11}
\end{figure}

The physical scenario becomes far more interesting if degeneracy of the $X_{\pm}$-points is established, which can be achieved by appropriate polarization management of the lattice beams as earlier mentioned. In this case the atomic sample has two equivalent minima for condensation and it needs to be clarified, which is the minimal energy state realized by the system. In Fig.~\ref{Fig.10}(d) momentum spectra for different settings of the condensation point energies $E(X_{\pm})$ are shown. The difference between $E(X_{\pm})$ in the left and right panels corresponds to about 0.5 nanokelvin. The spectrum in the center obviously shows that both condensation points are nearly equally populated. Each subsample in either of the points $X_{\pm}$ is coherent as demonstrated by the observation of sharp Bragg resonances, such that one may conclude that the atomic sample is a superposition of two condensates, which immediately raises the question, whether they possess a definite relative phase. Unfortunately, an unequivocal answer cannot be simply obtained from momentum spectra. To see this, let us describe the atomic sample as a superposition of condensates $|\psi_+\rangle$ and $|\psi_-\rangle$ associated with the $X_{\pm}$-points with a possibly indeterminate relative phase $\delta \phi$, i.e. $|\delta \phi, \pm \rangle \equiv |\psi_+\rangle \pm e^{i \delta \phi}|\psi_-\rangle$, where \textit{indeterminate} means that $\delta \phi$ may take arbitrary values in $[0,2\pi]$ in each experimental realization. Hence, the momentum spectrum is $|{\tilde \psi_{\delta \phi, \pm}}(p)|^2 = |{\tilde \psi}_+(p) \pm e^{i \delta \phi}{\tilde \psi}_-(p)|^2 = $ $|{\tilde \psi}_+(p)|^2+|{\tilde \psi}_-(p)|^2 \pm 2\cos(\delta \phi)\,{\tilde \psi}_+(p) \,{\tilde \psi}_-(p)$, where it has been assumed that $\psi_{\pm}(r) \equiv \langle r | \psi_{\pm} \rangle$ and thus the corresponding Fourier transforms ${\tilde \psi}_{\pm}(p)$ can be chosen real. Since ${\tilde \psi}_{\pm}(p)$ have disjoint supports, one gets ${\tilde \psi}_+(p)\, {\tilde \psi}_-(p) \approx 0$ and hence $|{\tilde \psi_{\delta \phi, \pm}}(p)|^2 \approx |{\tilde \psi}_+(p)|^2+|{\tilde \psi}_-(p)|^2$. As a consequence we cannot extract information on the relative phase $\delta \phi$ from momentum spectra. 

\subsection{Observing phase locking by matter wave interference}

Obtaining information on the relative phase $\delta \phi$ requires the use of interference techniques. A possible strategy has been proposed in Ref.~\cite{Li:14}. Before taking a momentum spectrum, a Bragg spectroscopy pulse \cite{Ste:99} is applied in order to mix the momentum space wave functions ${\tilde \psi}_{\pm}(p)$ by a two-photon coupling. If the two sub-states $|\psi_{\pm}\rangle$ in the condensation points $X_{\pm}$ possess a fixed relative phase, this leads to observable signatures in the subsequent time evolution. Here we focus on a second method that has been successfully implemented in Ref.~\cite{Koc:15}. The key idea is to split the tubular lattice sites into upper and lower sections by introducing a tunnel barrier in the $xy$-plane by means of a cylindrically focused blue-detuned laser beam, which exerts a repelling force upon the atoms. Thus, two identical perfectly aligned optical lattices arise below and above the tunnel barrier, which is chosen sufficiently high such that tunneling between both lattices is completely suppressed (see Fig.~\ref{Fig.11}(a)). Excitation of atoms into the second band and subsequent thermalisation is simultaneously and independently applied to both lattices. The initial BEC is positioned such that about half of the atoms are transferred to each side of the barrier. The lattice potential is adjusted to the case of degenerate condensation points $X_{\pm}$ such that equal populations of $X_{\pm}$ are prepared, as is confirmed by the band mapping spectrum in Fig.~\ref{Fig.11}(b) recorded with a line of sight parallel to the $z$-axis. Finally, ballistic expansion is applied during $t_{\textrm{TOF}}=25\,$ms until the atom clouds from each lattice overlap and interfere (see Fig.~\ref{Fig.11}(c)). 

\begin{figure}
\includegraphics[scale=0.5, angle=0, origin=c]{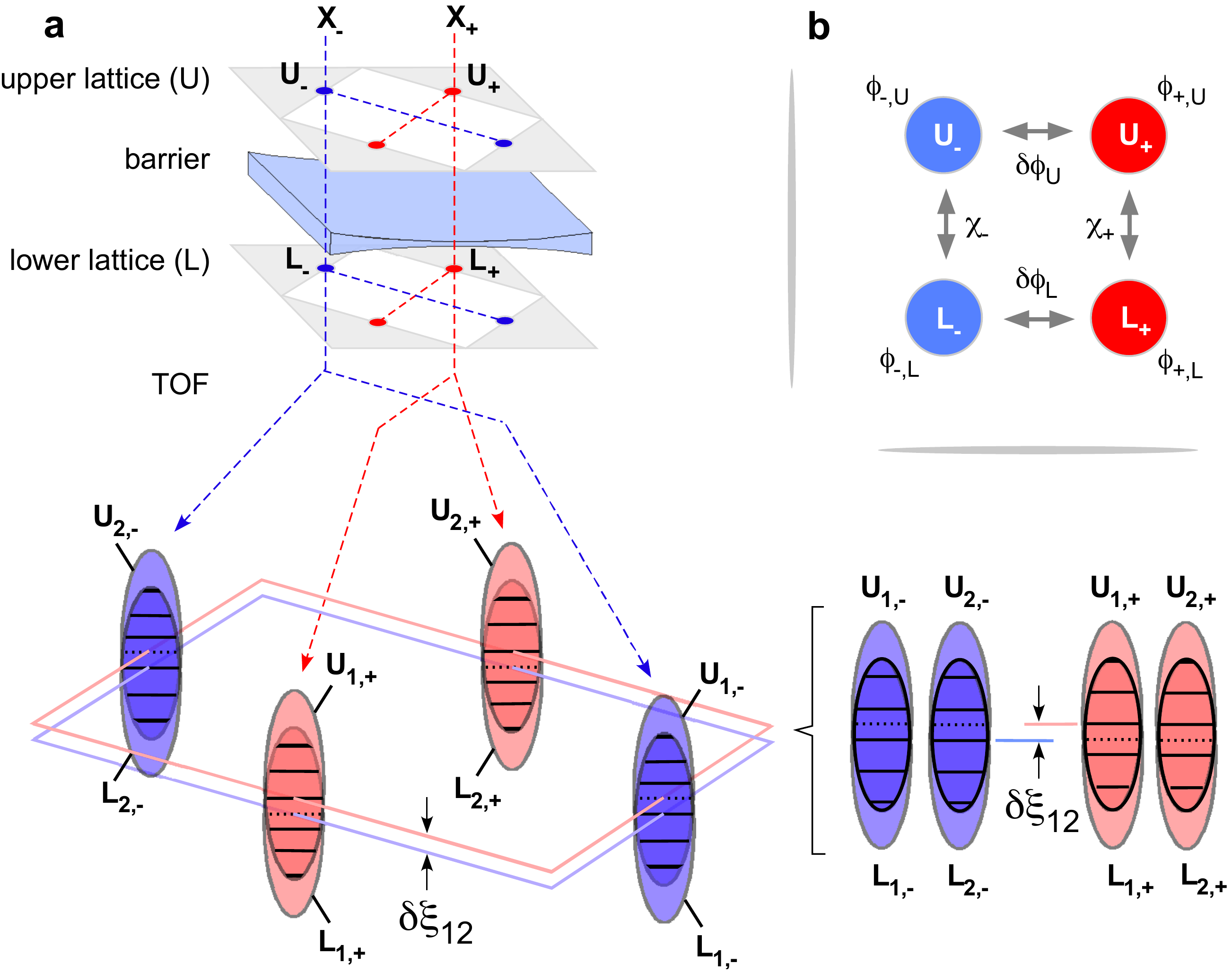}
\caption{(a) Condensates $U_{\sigma}$ and $L_{\sigma}$ are prepared in the condensation points $X_{\sigma}$ of the upper ($U$) and the lower ($L$) lattices ($\sigma \in \{-,+\}$). Ballistic expansion yields a decomposition of $U_{\sigma}$ and $L_{\sigma}$ into their two predominant momentum components $U_{i,\sigma}$ and $L_{i,\sigma}$ indexed $i \in \{1,2\}$ with the average momentum ${\bf p}_{i, \sigma} \equiv(-1)^i \frac{1}{2} \hbar k (\sigma \, 1,1)$. Ballistic expansion acts to overlap condensates $U_{\sigma}$ and $L_{\sigma}$ originating from the same $X_{\sigma}$-point, which thus interfere, while condensates from different $X_{\sigma}$-points are separated. This leads to the four density modulated Bragg maxima $B_{i, \sigma}$ of Fig.~\ref{Fig.11}(c) and (d). Density gratings $B_{i, \sigma}$ resulting from the same $X_{\sigma}$-point exhibit the same spatial offset phase with respect to the $xy$-plane. The relative spatial phase $\delta \xi_{ij}$ of density gratings $B_{i, \sigma}$ originating from different $X_{\sigma}$-points is indicated in the picture. (b) The red and blue disks symbolize the four condensates $U_{\sigma}$ and $L_{\sigma'}$ indicated in (a) with $\phi_{U,\pm}$, $\phi_{L,\pm}$ denoting their global phases. The grey arrows symbolize the relative phases $\chi_{\pm} \equiv \phi_{U,\pm} - \phi_{L,\pm}$,  $\delta \phi_L \equiv \phi_{-,L} - \phi_{+,L}$, $\delta \phi_U \equiv \phi_{U,-} - \phi_{U,+}$.}
\label{Fig.12}
\end{figure}

Similar to the case of a single lattice in Fig.~\ref{Fig.10}(d) (central panel), the density distribution in Fig.~\ref{Fig.11}(c) represents an image of momentum space. In the ballistic expansion, the sub-sample in each condensation point $X_{\sigma}$ is decomposed into its two predominant momentum components with regard to the $xy$-plane, thus giving rise to four zero-order Bragg peaks $B_{i, \sigma}$ (identified by the indices $({i, \sigma})$ in Fig.~\ref{Fig.11}(c)), where $\sigma \in \{-,+\}$ denotes the condensation point that $B_{i, \sigma}$ is associated with and $i \in \{1,2\}$ numbers the two predominant momentum classes ${\bf p}_{i, \sigma} \equiv(-1)^i \frac{1}{2} \hbar k (\sigma \, 1,1)$ associated with $X_{\sigma}$ (cf. Fig.~\ref{Fig.11}(b)). Since the momentum spectrum arises from the interference of two lattices closely spaced on top of each other, in contrast to the case of a single lattice, each Bragg peak exhibits a density modulation along the $z$-direction. In the simplified picture that the interference results from two point sources at a distance $d_z$ observed after sufficiently long time of flight $t_{\textrm{TOF}}$, the wavelength of the expected density grating is $\lambda_z = \frac{h\,t_{\textrm{TOF}}}{m\,d_z}$, with $h$ denoting Planck's constant and $m$ the atomic mass \cite{And:97}. This amounts to $\lambda_z \approx 10\,\mu$m, close to what is observed in the experiment. The density gratings $B_{i, \sigma}$ are recorded by absorption imaging with the line of sight indicated by the vector $\mathbf{r}$ in Fig.~\ref{Fig.11}(c), which lies in the $xy$-plane and encloses an angle of $\alpha=13^\circ$ with the $y$-axis. An example of a single-shot image of these density gratings is shown in Fig.~\ref{Fig.11}(d), where the new coordinates $\tilde{x},\tilde{z}$ are used for the projected density distribution in the image plane. As is illustrated in Fig.~\ref{Fig.12}(a), the density gratings $B_{i, \sigma}$ arise from the interference of sub-samples $U_{i, \sigma}$ and $L_{i, \sigma}$, which denote the components with average momenta ${\bf p}_{i, \sigma}$ of the condensates $U_{\sigma}$ and  $L_{\sigma}$ formed in the condensation points $X_{\sigma}$ of the upper ($U$) and lower ($L$) lattice, respectively. Note that only sub-samples $U_{\sigma}$ and $L_{\sigma}$ associated with the same condensation point $X_{\sigma}$ yield interference, while sub-samples from different condensation points are spatially separated in the ballistic expansion process.

The global spatial phases $\xi_{i,\sigma}$ of the gratings $B_{i, \sigma}$ vary for different experimental realizations, thus reflecting the complete independence of the atomic samples produced in the upper and lower lattice \cite{Jav:96}. Only relative spatial phases $\xi_{i,\sigma}-\xi_{j,\sigma'}$ may carry interesting information about the quantum state produced in both lattices. Since the sub-samples $U_{1, \sigma}$ and $U_{2, \sigma}$ and similarly $L_{1, \sigma}$ and $L_{2, \sigma}$ are associated with the same condensation point, they are expected to be phase coherent. Hence, density gratings $B_{i, \sigma}$ with different $i$ but the same $\sigma$ should exhibit zero relative spatial phase, i.e. $\xi_{1,\sigma} = \xi_{2,\sigma}$. To obtain relevant information we must consider the spatial phase difference $\delta \xi_{ij} \equiv \xi_{i,-} -\xi_{j,+}, $ of pairs of gratings $B_{i, -}$, $B_{j, +}$ associated with different condensation points. It turns out that $\delta \xi_{ij}$ in fact provides information on the phase relation between the condensates of different condensation points. To see this, consider the four condensates $U_{\pm}$ and $L_{\pm}$, symbolized by red and blue disks in Fig.~\ref{Fig.12}(b). Assume global phases $\phi_{U,\pm}$, $\phi_{L,\pm}$ of these condensates, which in the most general case may take arbitrary values in different experimental realizations via spontaneous breaking of U(1) symmetry. The associated mutual relative phases $\chi_{\pm} \equiv \phi_{U,\pm} - \phi_{L,\pm}$,  $\delta \phi_L \equiv \phi_{L,-} - \phi_{L,+}$, $\delta \phi_U \equiv \phi_{U,-} - \phi_{U,+}$ are indicated adjacent to the grey arrows. Recall that $B_{i, \sigma}$ arises from the interference of sub-samples $U_{i, \sigma}$ and $L_{i, \sigma}$ originating from the same condensation point. Hence, the spatial phases $\xi_{i,\sigma}$ of $B_{i, \sigma}$ are given by the phases $\chi_{\sigma}$ in Fig.~\ref{Fig.12}(b) plus some fixed geometric offset phase independent of $i$ and $\sigma$ with the consequence that $\delta \xi_{ij} = \chi_{-} - \chi_{+} =  \delta \phi_U - \delta \phi_L$, irrespective of the choice of $i,j \in \{1,2\}$. Therefore, by measuring $\delta \xi_{ij}$ we may determine the difference between the relative condensate phases in the upper and lower lattices.

\begin{figure}
\includegraphics[scale=0.6, angle=0, origin=c]{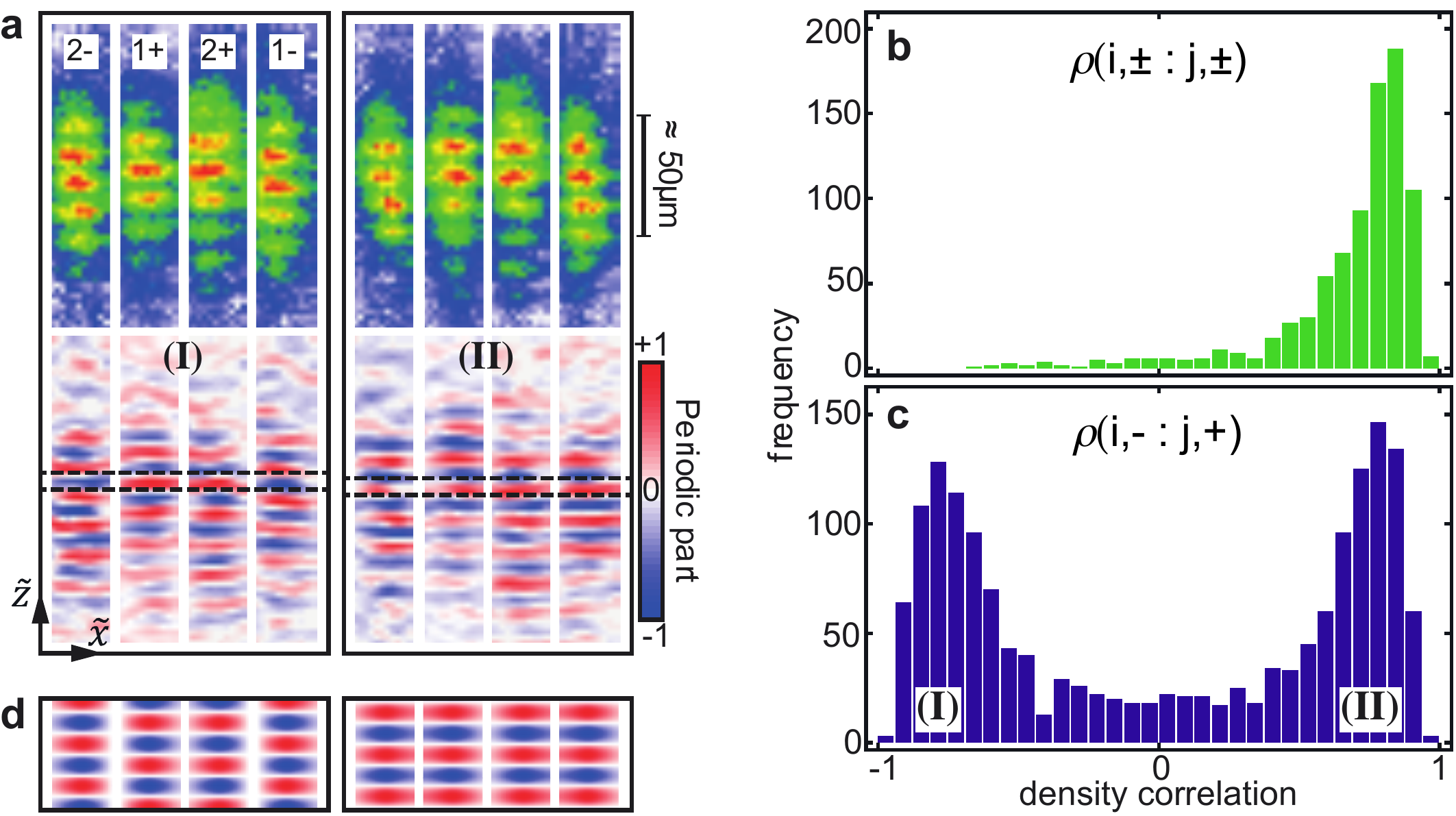}
\caption{(a) The left (I) and right (II) panels show examples of the two classes of observed interference patterns. The upper row shows the raw data, which are repeated in the lower row after band-pass filtering and normalization. A data set comprises four density gratings $n_{i,\pm}(\tilde{z})$ with $i\in \{1,2\}$ corresponding to the four Bragg maxima in Fig.~\ref{Fig.11}(c) and (d) (see text for definition of $n_{i,\pm}(\tilde{z})$). The black dashed boxes highlight the spatial correlations among the four density gratings. (b) Stacked histogram for $\rho(i,\sigma : j,\sigma')$ with $i \neq j \in \{1,2\}$ and $\sigma=\sigma'\in \{-,+\}$, which correlate different density gratings associated with the same condensation point. (c) Stacked histogram for the density-density cross correlations $\rho(i,- : j,+)$ for different density gratings associated with different condensation points. (d) Calculated interference confirming the spatial correlations found in (a). Reproduced from Ref.~\cite{Koc:15}, \copyright\ American Physical Society.}
\label{Fig.13}
\end{figure}

The results of a correlation analysis of the observed interference gratings are summarized in Fig.~\ref{Fig.13}. A single-shot absorption image of all four Bragg peaks (in the ($\tilde{x}$,$\tilde{z}$)-plane) is shown in Fig.~\ref{Fig.13}(a). Since one is only interested in the spatial phases of the periodic parts of the interference gratings along the $\tilde{z}$-direction, the constant offset is removed by band-pass filtering thus obtaining the normalized distributions $n_{i,\pm}(\tilde{x},\tilde{z})$ shown in the lower row of Fig.~\ref{Fig.13}(a). These distributions take values in the interval $[-1,1]$  (blue to red) and are centered around zero (white): $\sum_{\tilde{x},\tilde{z}} n_{i,\pm}(\tilde{x},\tilde{z})\approx 0$. After summing along the pixel values in the $\tilde{x}$-direction in order to obtain $n_{i,\pm}(\tilde{z}) \equiv \sum_{\tilde{x}} n_{i,\pm}(\tilde{x},\tilde{z})$, one may evaluate the normalized density correlations
\begin{eqnarray}
\rho(i,\sigma : j,\sigma') \equiv \frac{\sum_{\tilde{z}} n_{i,\sigma}(\tilde{z})n_{j,\sigma'}(\tilde{z})}{\sqrt{\sum_{\tilde{z}} n_{i,\sigma}^2(\tilde{z}) \sum_{\tilde{z}} n_{j,\sigma'}^2(\tilde{z})}} , \,\,\,  \sigma, \sigma' \in\{-,+\}.
\label{Eq.3}
\end{eqnarray}
The correlations $\rho(i,\sigma : j,\sigma')$ take positive values near $+1$ for in-phase density gratings ($\xi_{i,\sigma}-\xi_{j,\sigma'}=0$) and negative values near $-1$ for out-of-phase gratings ($\xi_{i,\sigma}-\xi_{j,\sigma'}=\pi$). Fig.~\ref{Fig.13}(b) shows a histogram for the observed values of $\rho(i,\sigma : j,\sigma')$ with $i \neq j \in \{1,2\}$ and $\sigma=\sigma'\in \{-,+\}$, which correlate different density gratings associated with the same condensation point. A single accumulation point is observed near $+1$, in accordance with the expectation that Bragg peaks resulting from equivalent condensation points of the upper and lower lattices should carry in-phase interference gratings. A far more surprising results is shown in Fig.~\ref{Fig.13}(c), which displays the histogram for the correlations $\rho(i,- : j,+)$ with $i,j \in \{1,2\}$ for density gratings associated with different condensation points. In contrast to (b), a bimodal distribution with practically equal numbers of correlated and anti-correlated patterns is observed. This shows that $\delta \xi_{ij} \in \{0,\pi \}$ and hence $\delta \phi_u - \delta \phi_l \in \{0,\pi \}$, i.e. the phase differences $\delta \phi_u$ between the condensates $U_{+}$ and $U_{-}$ and $\delta \phi_l$ between $L_{+}$ and $L_{-}$ are locked. Since the condensates in the upper and lower lattices are completely independent with no phase relation established, this can only hold if independently $\delta \phi_u \in \{ \delta \phi, \delta \phi + \pi \}$ and $\delta \phi_l \in \{ \delta \phi, \delta \phi + \pi \}$ with the same inherently fixed phase $0 \leq \delta \phi < \pi$, which is the same in each experimental implementation. In other words, in each lattice a single condensate is formed in either of the two possible states $|\delta \phi, \pm\rangle \equiv |\psi_+\rangle \pm e^{i \delta \phi}|\psi_-\rangle$ with a fixed phase $\delta \phi$ inherent to the system. The different signs $\pm$ occur with equal probability in each experimental implementation, reflecting a spontaneously broken $\mathbb{Z}_2$ symmetry. Although this unequivocally demonstrates phase locking of the subsamples condensed in different condensation points, the value of $\delta \phi$ is yet to be determined. This is the topic of the following subsection. We note here that only at the lowest temperatures the observed interference pattern exhibit the simple form discussed here. At higher temperatures excitations yield interference patterns with additional nodal structures, which for example can indicate the formation of domains with different values of $\delta \phi$ (See Ref.~\cite{Koc:15} for a detailed discussion).

\begin{figure}
\includegraphics[scale=0.5, angle=0, origin=c]{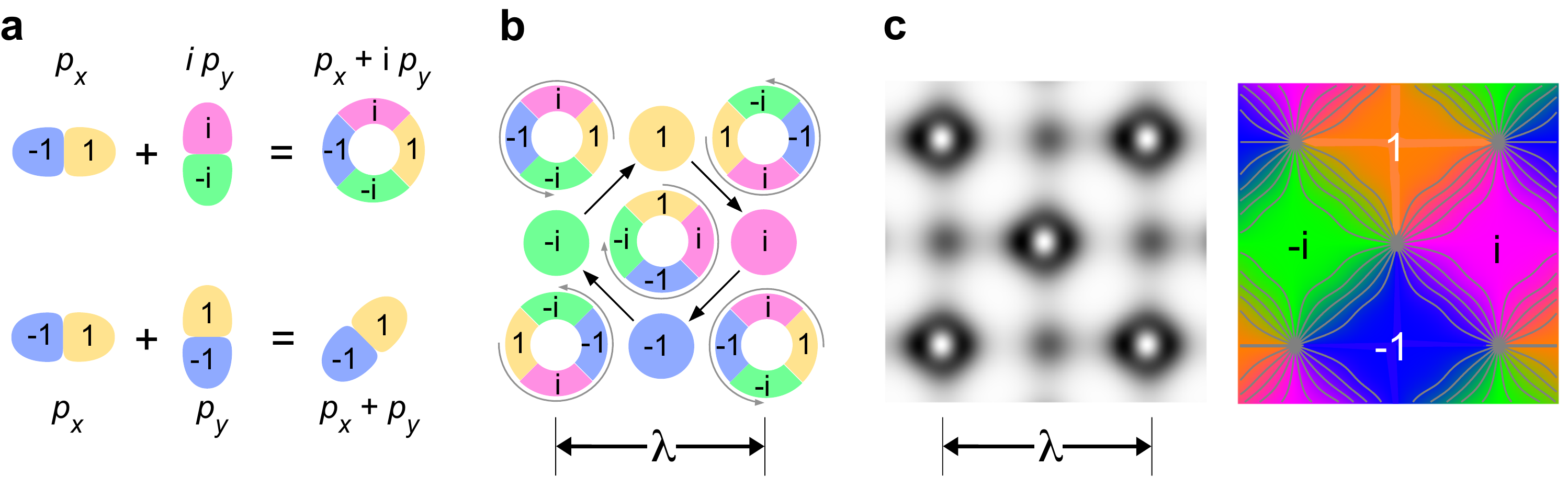}
\caption{(a) Superpositions of $p_x$- and $p_y$-orbitals with maximal (upper row) and minimal (lower row) mode volume. (b) Schematic of the wave functions associated with the states $|\pi/2,\pm \rangle$. (c) The left and right panels show the probability density $|\psi_{+} + i \psi_{-}|^2$ (dark regions correspond to high probability)) and the complex argument $\textrm{Arg}(\psi_{+} + i \psi_{-})$, respectively, calculated for $V_{xy,0} = 7\,E_{\textrm{rec}}$ and $\theta = 0.75\,\pi$. The colors orange, magenta, blue, green correspond to phase values $1, i, -1, -i$, respectively.}
\label{Fig.14}
\end{figure}

\subsection{Determination of the relative phase between the condensates in $X_{\pm}$}

Knowing that the relative phase between the condensates in $X_{\pm}$ is locked we may determine its value $\delta \phi$ by minimizing the free energy for the family of states $|\delta \phi, \pm\rangle$. These states are superpositions of $|\psi_{\pm}\rangle$, which are degenerate with respect to their single-particle energy. Hence, $|\delta \phi, \pm\rangle$ exhibit the same single-particle energy for all values of $\delta \phi$. However, as has been discussed in Refs.~\cite{Isa:05, Liu:06, Wu:09} for a monopartite square lattice, in case of an arbitrarily small contact interaction via binary on-site collisions, the total energy can acquire a $\delta \phi$ dependence. Within the local $s$-orbitals in the shallow wells both Bloch functions $\langle r |\psi_{\pm}\rangle$ are approximated by the same wave function $s(r)$ such that $\langle r |\delta \phi, \pm \rangle  \approx$ $s(r) \pm e^{i \delta \phi} s(r)$. In the deep wells $\langle r |\psi_{\pm}\rangle \approx p_{\pm}(r)$, where $p_{\pm}(r) \equiv p_{x \pm y}(r) = (p_{x}(r)\pm p_{y}(r))/\sqrt{2}$ denote the wave functions of the local $p$-orbitals aligned along the $x\pm y$-directions. Therefore, in the deep wells $\langle r |\delta \phi, \pm \rangle \approx p_{+}(r) \pm e^{i \delta \phi} \,p_{-}(r)$. In the evaluation of the interaction energy only the terms associated with $p$-orbitals turn out to be relevant. The local interaction integral $U(\delta \phi) \equiv g  \int d^2r |\langle r |\delta \phi, \pm \rangle|^4$, for repulsive interaction ($g > 0$), takes its maximum at $\delta \phi = 0$ and its minimum at $\delta \phi = \pi/2$. The underlying physics resembles Hund's second rule in multi-electron atoms: the interaction energy is minimized if $\delta \phi = \pi/2$, because the superposition $p_{x} \pm i \, p_{y}$ provides the largest possible mode volume such that the repulsively interacting atoms can best avoid each other (See Fig.~\ref{Fig.14}(a)). Due to the degeneracy of the band minima, an arbitrarily small interaction energy can thus determine which quantum state is realized. Assuming that the deep wells host $p_{+} \pm i \, p_{-}$-orbitals and the shallow wells host $s$-orbitals, we may compose the global wave function via the rule that minimization of kinetic energy requires equal local phases at the tunneling bonds. This leads to the schematic wave function shown in Fig.~\ref{Fig.14}(b), which nicely reproduces the geometry of the wave functions $\langle r | \frac{\pi}{2}, \pm \rangle$ determined in a numerical band calculation. This is illustrated in Fig.~\ref{Fig.14}(c), which shows the probability density and the complex argument of $\langle r | \frac{\pi}{2}, + \rangle$ in the left and right panel, respectively, calculated for the potential in Eq.~\ref{Eq.1} with $V_{xy,0} = 7\,E_{\textrm{rec}}$, $\theta = 0.75\,\pi$ and the distortion parameters $\eta = 1.03$, $\epsilon_{x} = 0.93$, and $\epsilon_{y} = 1$. The local phases are arranged to introduce orbital currents and plaquette currents breaking time-reversal symmetry and the translation symmetry of the lattice. Note also the $\mathbb{Z}_2$ symmetry associated with the two possible signs of $\langle r | \frac{\pi}{2}, \pm \rangle$, which is spontaneously broken in experimental observations. 

\begin{figure}
\includegraphics[scale=0.5, angle=0, origin=c]{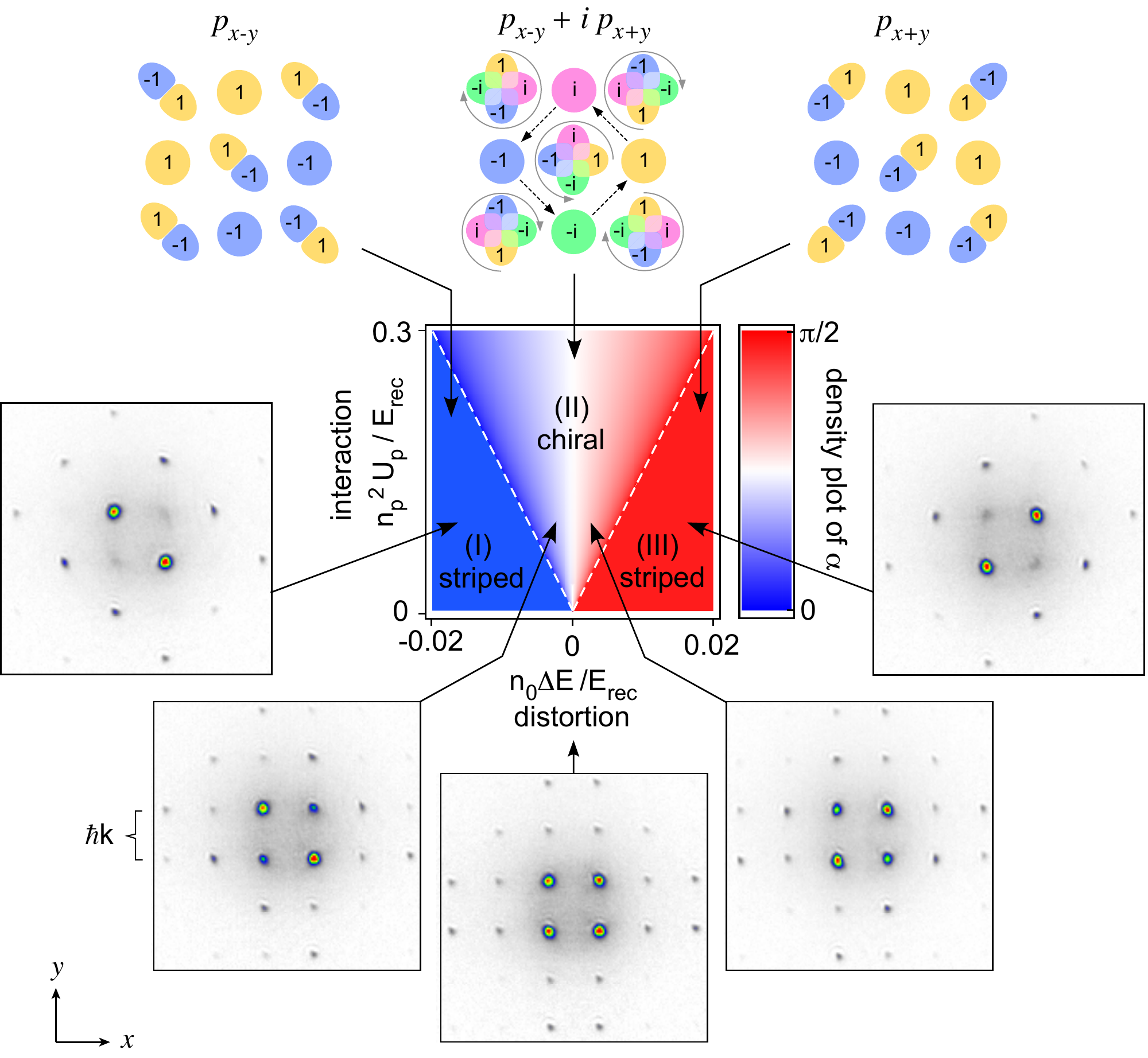}
\caption{The phase diagram in the centre shows the mixing angle $\alpha$ in Eq.~\ref{Eq.6} versus $n_{p}^2\,U_{p}$ and $n_0 \, \Delta E$ in units of $E_{\textrm{rec}}$. Three phases arise (regions (I),(II), and (III)) separated by 2nd-order transitions (white dashed lines). In (I) and (III) only a single condensation point is populated such that striped order prevails. In the central region (II) both condensation points are populated and chiral order emerges. Reproduced from Ref.~\cite{Oel:13}.}
\label{Fig.15} 
\end{figure}

The preceding discussion can be made more quantitative by the following simple mean-field consideration. Let us consider the family of wave functions $\psi(r) \equiv \cos(\alpha)\,\psi_{+}(r) + \sin(\alpha) \,e^{i \delta \phi} \psi_{-}(r)$ describing an arbitrary superposition of the Bloch functions $\psi_{\pm}(r)$ associated with the condensation points $X_{\pm}$. Assume that $\psi_{\pm}(r)$ are real eigenfunctions of the single-particle Hamiltonian $H$ with $H \psi_{\pm} = \pm \frac{1}{2} \Delta E\,\psi_{\pm}$ with $\Delta E \equiv E(X_+)-E(X_-)$. Normalizing to $n_0$ particles per unit cell "$\diamond$" requires $n_0 = \int_{\diamond} d^2r |\psi_{\pm}|^2$ and thus $n_0 = \int_{\diamond} d^2r |\psi|^2$. We now consider the total energy functional $E[\alpha,\delta \phi] \equiv \int_{\diamond} d^2r \,\left( \psi^{*} H \psi + \frac{g}{2} |\psi|^4 \right)$ where $g>0$ is a positive constant quantifying the repulsive contact interaction. Defining the overlap integrals $\rho_0 \equiv \int_{\diamond} d^2r\, \psi_{\pm}^4$, $\rho_1 \equiv \int_{\diamond} d^2r\, \psi_{+}^2\psi_{-}^2$ and $\rho_2 \equiv \int_{\diamond} d^2r\, \psi_{\mp} \psi_{\pm}^3$ one finds
\begin{eqnarray}
\label{Eq.4}
E[\alpha,\delta \phi]  &=&  n_0 \,\frac{1}{2} \Delta E \left[ \sin^2(\alpha) - \cos^2(\alpha) \right] 
\\ \nonumber 
&+& \frac{g}{2} \rho_0 \left[ \sin^4(\alpha) + \cos^4(\alpha) \right] + g \rho_1 \left[1 + 2 \cos^2(\delta \phi) \right] \sin^2(\alpha) \cos^2(\alpha) + 2 g \rho_2 \sin(\alpha) \cos(\alpha) \cos(\delta \phi) \,\,.
\end{eqnarray}
\label{Eq.5}
After setting $\rho_2=0$ for symmetry reasons, minimization of $E[\alpha,\delta \phi]$ with respect to $\alpha$ and $\delta \phi$ yields 
\begin{eqnarray}
\cos(2\alpha) = -\frac{n_0\,\Delta E}{g\, (\rho_0 - \rho_1)} \,\, \,\, \textrm{and}  \,\,\,\, \delta \phi = \frac{\pi}{2}\,.
\end{eqnarray}
According to Fig.~\ref{Fig.10}(b), we now specify the Bloch functions $\psi_{\pm}(r)$ in a typical plaquette to be composed of local $s(r)$-orbitals in the shallow wells and local $p_{\pm}(r)$-orbitals in the deep wells, i.e. $\psi_{\pm}(r) = \sqrt{n_s}\,s(r) + \sqrt{n_p}\,p_{\pm}(r)$ with normalized wave functions $1=\int_{\diamond} d^2r |s(r)|^2 = \int_{\diamond} d^2r |p_{\pm}(r)|^2$. The fractional populations $n_s$ and $n_p$ of the shallow and deep wells, respectively, satisfy $n_s + n_p =n_0$. Assuming that $s(r)$ and $p_{\pm}(r)$ have approximately disjoint supports, one finds $g\,(\rho_0 - \rho_1) = \frac{g}{2} n_p^2 \int_\diamond d^2r (|p_{+}(r)|^2-|p_{-}(r)|^2)^2$. In the harmonic approximation this yields $g\,(\rho_0 - \rho_1) = \frac{3}{2} n_p^2 \, U_p$ with the interaction per particle in the $p$-orbitals $U_p \equiv g \int_\diamond d^2r |p_{\pm}(r)|^4$ and thus
\begin{eqnarray}
\cos(2\alpha) = \frac{3 \,n_0  \, \Delta E}{2 \,n_p^2 \,U_p} \,.
\label{Eq.6}
\end{eqnarray}
This result agrees with a more involved calculation assuming a three-band tight-binding model reported in Ref.~\cite{Oel:13}. A numerical mean-field calculation based upon the Gross-Pitaevskii equation is found in Ref.~\cite{Cai:11}. Eq.~\ref{Eq.6} gives rise to the phase diagram plotted in Fig.~\ref{Fig.15} with respect to the two experimentally tunable parameters $n_0\, \Delta E$ and $n_p^2 \,U_p$. The latter can be readily tuned by varying $n_p$ via adjustment of $\theta$ (cf. Eq.~\ref{Eq.1}), even if $U_p$ is fixed. Three phases arise (regions (I),(II), and (III)), separated by 2nd order transitions (white dashed lines). In (I) and (III) only a single condensation point is populated such that striped order prevails. In the central region (II) both condensation points are populated and the wave function displays chiral order as already discussed in Fig.~\ref{Fig.14}(b). 

\begin{figure}
\includegraphics[scale=0.5, angle=0, origin=c]{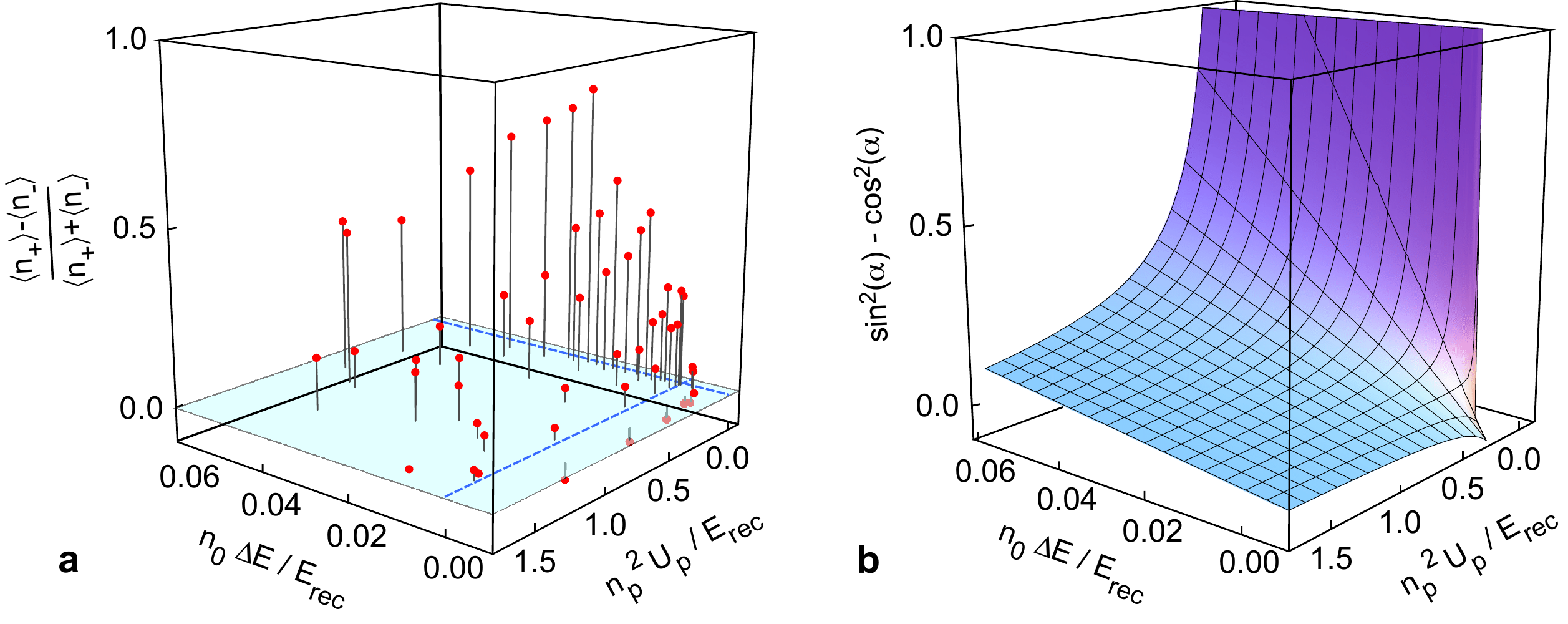}
\caption{(a) Plot of the relative population difference $\langle n_{+} - n_{-}\rangle /\langle n_{+} + n_{-} \rangle$ of the condensation points versus the parameters tuned in the phase diagram in Fig.~\ref{Fig.15}. The brackets denote the average over five experimental measurements. (b) Calculation for a homogeneous infinite sample according to Eq.~\ref{Eq.6}.}
\label{Fig.16} 
\end{figure}

\begin{figure}
\includegraphics[scale=0.5, angle=0, origin=c]{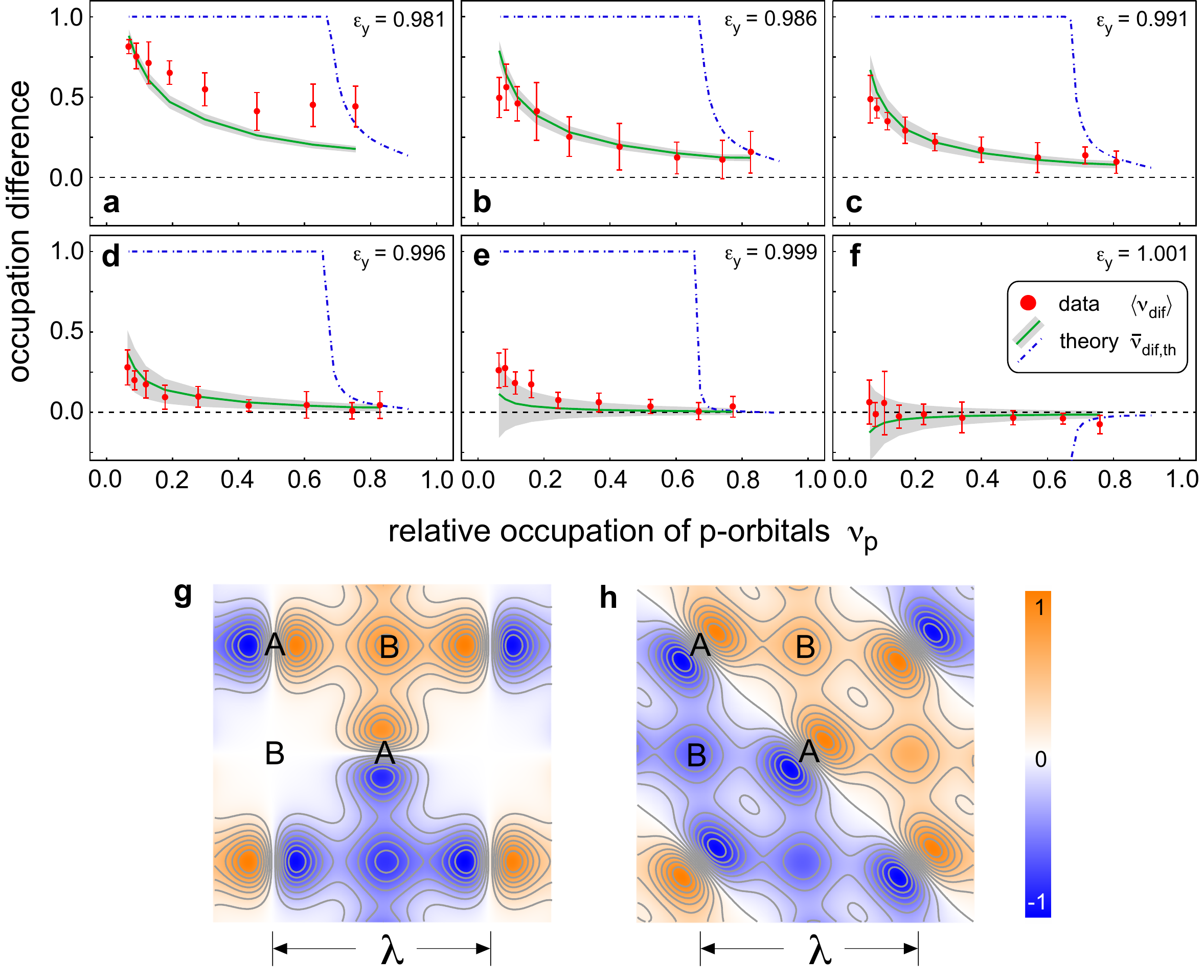}
\caption{The normalized mean occupation difference $\langle \nu_{\text{dif}}\rangle = \langle n_{+} - n_{-}\rangle /\langle n_{+} + n_{-} \rangle$ between the two condensation points $X_{\pm}$ is plotted versus the relative occupation of the $p$-orbitals $\nu_{p} =n_{p}/n_{0}$. The error bars show the statistical errors for eight measurements. The adjusted distortion of the lattice potential is $1-\epsilon_y = 0.019, 0.014, 0.009, 0.004, 0.001, -0.001$ from (a) to (f). The solid green lines show the corresponding theoretical predictions $\overline{\nu}_\text{dif,th} = \sin^2(\alpha)-\cos^2(\alpha)$ derived by means of Eq.~\ref{Eq.6}. The grey areas represent the uncertainty of the lattice distortion $\Delta\epsilon_y=\pm 2.5 \cdot 10^{-3}$. The blue dashed lines show the predictions for an incoherent mixture of condensates at $X_{\pm}$. In (g) and (h) the real superpositions $\cos(\alpha)\,\psi_{+}(r) + \sin(\alpha) \, \psi_{-}(r)$ are shown with $\alpha = \pi/4$ in (g) and $\alpha = 0$ in (h). The positions of the deep and shallow wells are marked by $\mathcal{A}$ and  $\mathcal{B}$, respectively. Parts of the figure are taken from Ref.~\cite{Oel:13}.}
\label{Fig.17} 
\end{figure}

The phase diagram shows that for small values of the interaction in the $p$-orbitals a small distortion of the $C_4$ symmetry of the lattice can yield a significant depletion of one of the condensation points $X_{\pm}$, while at large values of the  interaction the system prefers equilibrated populations of $X_{\pm}$. This property of the system has been directly tested experimentally in Ref.~\cite{Oel:13}. In Fig.~\ref{Fig.16}(a) the relative population difference $\langle n_{+} - n_{-}\rangle /\langle n_{+} + n_{-} \rangle$ of the condensation points $X_{\pm}$, averaged for five measurements, is plotted versus the lattice distortion parameter $n_0 \, \Delta E$ and the interaction parameter $n_p^2 \,U_p$ tuned in the phase diagram of Fig.~\ref{Fig.15}. The expected behavior is observed: If $n_p^2 \,U_p$ is small, a small deviation of $n_0 \, \Delta E$ from zero in fact yields a significant difference between $n_{+}$ and $n_{-}$. For large $n_p^2 \,U_p$ the difference $n_{+} - n_{-}$ remains close to zero even for large lattice distortions $n_0 \, \Delta E$. In Fig.~\ref{Fig.16}(b) the relative population difference $\overline{\nu}_\text{dif,th} = \sin^2(\alpha)-\cos^2(\alpha)$ calculated from Eq.~\ref{Eq.6} is shown, which agrees with the experimental findings in (a). In a finite atomic sample confined in a trap potential applied in addition to the lattice, the population per plaquette $n_0$ and hence the interaction parameter $n_p^2 \,U_p$ necessarily varies across the trap such that a shell structure is expected to occur with a chiral inner part of the lattice but striped regions at its boundary. This has been accounted for in Ref.~\cite{Oel:13} and quantitative agreement between experiment and theory has been obtained as is shown in Fig.~\ref{Fig.17}. In this plot $n_p^2\,U_p$ is tuned for various values of $n_0 \Delta E$ parametrized by the distortion parameter $\epsilon_y$ (see Eq.~\ref{Eq.1}). The dependence of $\Delta E$ upon $\epsilon_y$ is obtained from a numerical band calculation using the potential in Eq.~\ref{Eq.1}. $\epsilon_y =1$ corresponds to $\Delta E=0$. 

Chirality emerges here as a way of the system to minimize interaction energy. The corresponding theoretical predictions in Fig.~\ref{Fig.15} and Fig.~\ref{Fig.16}(b) are nicely confirmed by the observations in Fig.~\ref{Fig.16}(a) and Fig.~\ref{Fig.17}. Without chirality, i.e. $\delta \phi = 0$, the theoretical predictions are not compatible with the observations: Minimizing $E[\alpha,\delta \phi]$ in Eq.~\ref{Eq.4} with respect to $\alpha$ for $\delta \phi = 0$ yields the condition $ \sin(2\alpha)[n_0 \, \Delta E+g(3\rho_1-\rho_0)\cos(2\alpha)]+g\rho_2\cos(2\alpha) =0$. Since for symmetry reasons $\rho_2$ is very small, the approximate solution is $\sin(2\alpha)\approx 0$ and hence $\alpha \in \{0,\pi/2\}$ or $\overline{\nu}_\text{dif,th} = \sin^2(\alpha)-\cos^2(\alpha) = \pm 1$. Consequently, the energetically favorable non-chiral wave functions are those associated with occupation of a single condensation point. A common occupation of both condensation points is predicted not to occur for $\delta \phi = 0$. This in sharp contrast with the experimentally observed emergence of equal populations of $X_{+}$ and $X_{-}$ upon increasing $n_p$. This energy argument is readily understood by comparing the real superpositions $\cos(\alpha)\,\psi_{+}(r) + \sin(\alpha) \, \psi_{-}(r)$ with equal populations of both condensation points $X_{\pm}$ (i.e., $\alpha=\pi/4$) or with only a single condensation point ($X_+$) occupied (i.e., $\alpha=0$). In the former case ($\alpha=\pi/4$), plotted in Fig.~\ref{Fig.17}(g), destructive interference eliminates the $s$-orbital occupation in every second shallow $\mathcal{B}$-well. In contrast, the wave function corresponding to $\alpha=0$, plotted in Fig.~\ref{Fig.17}(h), permits population of the $s$-orbitals in all $\mathcal{B}$-wells. In the deep $\mathcal{A}$-wells both wave functions only differ with respect to the orientation of the $p$-orbitals, which is energetically irrelevant. Hence, for real superpositions of $\psi_{+}(r)$ and $\psi_{-}(r)$, irrespective of the value of $n_p$, occupation of a single condensation point is always favored energetically, because this permits more space for the atoms to avoid each other, thus minimizing collisional interactions. 

The scenario of an incoherent mixture of spatially superimposed condensates at the $X_{\pm}$-points, which has been already ruled out by the interference experiment of Fig.~\ref{Fig.11}, may also be excluded by energy arguments. For this case we minimize the total energy $2\pi \, E[\alpha] \equiv \int_{[0,2\pi]}d\delta\phi \,E[\alpha,\delta\phi]$ with respect to $\alpha$ with $E[\alpha,\delta\phi]$ according to Eq.~\ref{Eq.4}. One finds that values of $\nu_{\text{dif,th}}$ deviating from $\pm 1$ require $\nu_{p} > 2/3$ and satisfy
\begin{eqnarray}
\label{mixture}
\nu_{\text{dif,th}} =  \frac{3 \, \Delta E}{[\nu_{p}^2-4\,(1-\nu_{p})^2]\,n_{0}\,U_{p}} \,.
\end{eqnarray}
An average over all plaquettes of the lattice yields the blue dashed traces in Fig.~\ref{Fig.17}, which clearly disagree with the observations.

\begin{figure}
\includegraphics[scale=0.5, angle=0, origin=c]{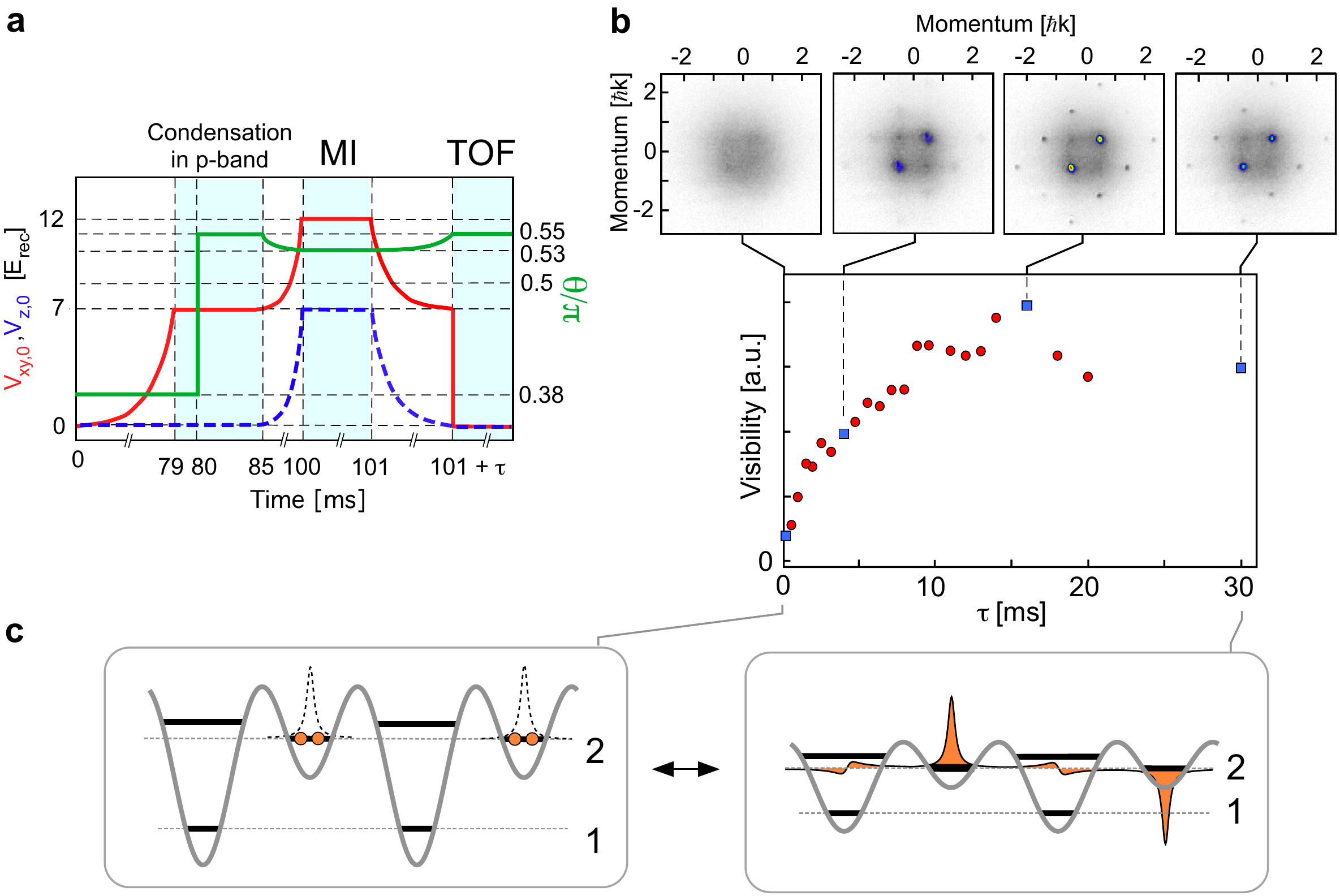}
\caption{(a) Experimental protocol for tuning of $V_{xy,0}$ (red solid line), $V_{z,0}$ (blue dashed line), and $\theta$ (green solid line). MI indicates the presence of a Mott insulator. TOF indicates the time interval, where a momentum spectrum is recorded via the time-of-flight technique. (b) Visibility plotted versus rephasing time $\tau$. Momentum spectra corresponding to the data points indicated by blue squares are shown on the upper edge. (c) Schematic of the Mott-insulator state corresponding to $\tau = 0\,$ms (left panel) and the superfluid state restored at $\tau = 30\,$ms. The atoms in the Mott state exclusively occupy the local $s$-orbitals of the shallow wells.}
\label{Fig.18} 
\end{figure}

\subsection{$p$-band Mott insulators}

Even in a simple square lattice potential with a monopartite unit cell the second band provides a rich structure of different Mott-insulator states, depending on the occupation \cite{Cha:09, Li:11, Col:10, Heb:13}. For unity occupation the inclusion of super-exchange interactions is predicted to energetically favor alternating $p_x$/$p_y$ orbital order \cite{Li:11}, while for occupations above unity staggered angular momentum ordering  is found \cite{Heb:13}. Further complexity arises in the chequerboard lattice considered in our experiments. For example, it is predicted in Ref.~\cite{Mar:12} that a nonzero condensate order parameter in the $\mathcal{A}$ sublattice can coexist with a Mott phase in the $\mathcal{B}$ sublattice. In experiments, the observation of Mott insulator physics is significantly hampered by band relaxation. Occupations of the deep wells exceeding unity yield fast collisional loss dynamics, if the lattice potential is increased to values suitable for the emergence of Mott insulating states. Careful design of the lattice potential giving rise to an increased anharmonicity of the wells might improve the situation in future experiments. Protection from band relaxation also arises, if only the shallow wells are notably occupied, which can be realized by choosing an appropriate value of $\Delta V$, as illustrated in Fig.~\ref{Fig.4}(b). If the lattice depth is adiabatically increased for this specific situation, a Mott insulator can be formed with all atoms populating local $s$-orbitals in the shallow wells, while the $p$-orbitals in the deep wells remain empty. This trivial Mott state is sufficiently long-lived to be observed in present experiments.

To address this Mott-state experimentally we have proceeded from a slightly distorted lattice ($\Delta E < 0$) such that a condensate is produced in the $X_{+}$ point according to the striped phase sector (III) of the phase diagram in Fig.~\ref{Fig.15}. The average well depth was ramped up in about 80~ms to $V_{xy,0} = 7\,E_{\textrm{rec}}$ with the external harmonic trap providing confinement in the $z$-direction ($V_{z,0}=0$), thus loading a BEC into the lowest band. By switching $\theta$ from about $0.38 \pi$ to $0.55\,\pi$, the atoms were swapped into the second band, with the band lifetime adjusted nearly to its maximum according to Fig.~\ref{Fig.4}(c). Subsequently, in 5~ms $V_{xy,0}$ and $V_{z,0}$ are adiabatically ramped up to $12\,E_{\textrm{rec}}$ and $7\,E_{\textrm{rec}}$, respectively, with an exponential time dependance. To keep the $p$-orbitals at minimal occupation, $\theta$ is meanwhile slightly adjusted to $\approx 0.53\,\pi$. After a fixed waiting time of 1~ms $V_{xy,0}$ and $V_{z,0}$ are exponentially reduced to $7\,E_{\textrm{rec}}$ and zero, respectively, during variable time durations $\tau$, while $\theta$ is tuned back to $\approx 0.55\,\pi$. The timing is summarized in Fig.~\ref{Fig.18}(a). In Fig.~\ref{Fig.18}(b), the visibility (red disks) of the $(1,1) \, \hbar k / 2$  Bragg resonance is plotted versus $\tau$. Examples of momentum spectra, corresponding to the data points indicated by blue squares, are shown by the insets at the upper edge of the graph. For $\tau = 0$ the visibility vanishes, indicating full loss of coherence as expected for a Mott insulator. As $\tau$ increases, phase coherence is restored and the visibility of the Bragg resonances revives. In Fig.~\ref{Fig.18}(c) the lattice potential and the occupations of the deep and shallow wells are sketched for the Mott state at $\tau = 0\,$ms and the superfluid state at $\tau = 30\,$ms. Although the Mott state (left panel) is melted to form a superfluid state belonging to the $p$-band (right panel), it is merely comprised of occupied $s$-orbitals and hence does not provide interesting orbital physics.

\begin{figure}
\includegraphics[scale=0.8, angle=0, origin=c]{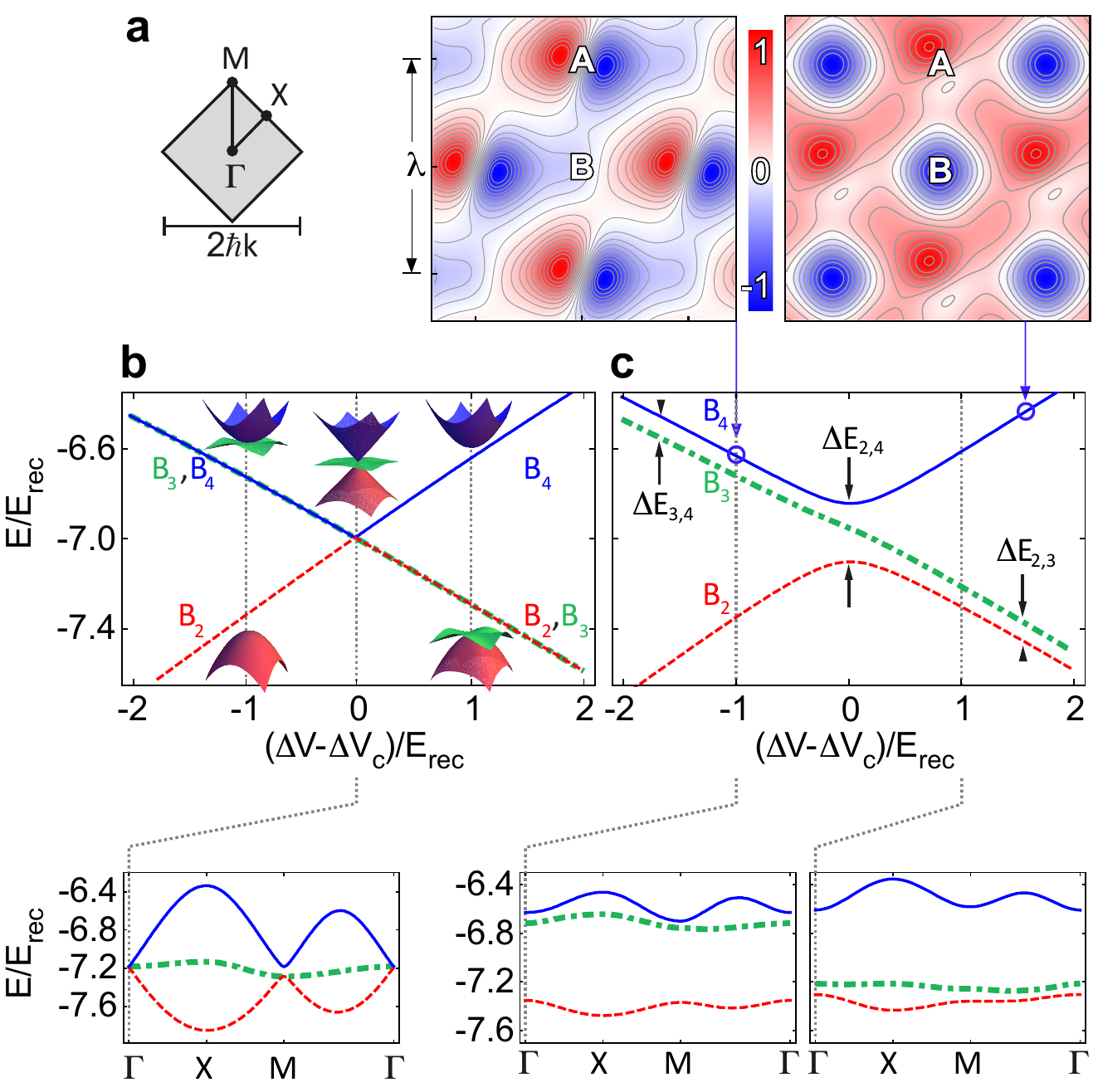}
\caption{(a) The first BZ with a path marked connecting the $\Gamma$-, $X$- and $M$-points. (b) Energies of the 2nd ($B_2$), 3rd ($B_3$), and 4th ($B_4$) Bloch bands at the $\Gamma$-point plotted versus $\Delta V$ for a lattice with perfect $C_4$-symmetry for $V_{xy,0} = 7.0 E_{\textrm{rec}}$ and $\Delta V_{c} = 6.1 \,E_{\textrm{rec}}$. The inset on the lower edge shows a plot of these bands for $\Delta V = \Delta V_{c}$ within the 1st BZ along the trajectory illustrated in (a). (c) The bands of (b) are replotted for a lattice with weakly broken $C_4$-symmetry for $V_{xy,0} = 7.8\,E_{\textrm{rec}}$ and $\Delta V_{c} = 6.1  \, E_{\textrm{rec}}$. The insets on the lower edge show plots analogue to that shown in (b), however, for $\Delta V = \Delta V_{c} \pm E_{\textrm{rec}}$. The insets on the upper edge show the Bloch functions at the $\Gamma$-point at the values of $\Delta V$ indicated by the blue circles on the left and the right of the avoided band crossing. Red and blue color correspond to positive and negative values, respectively. Nodal points appear in white. Parts of the figure are taken from Ref.~\cite{Oel:12}, \copyright\ American Physical Society.}
\label{Fig.19} 
\end{figure}

\section{Topological features and condensate dynamics in the fourth band}

Topologically non-trivial band structures are at the basis of intriguing forms of quantum matter such as high-temperature superconductors, unconventional superconductors in heavy-fermion compounds, quantum Hall systems or the recently discovered topological semi-metals in semiconductors with strong spin-orbit coupling \cite{Tok:00, Mae:04, Sto:99, Hsi:08, Has:10, Qi:11}. The unique possibilities to form tailor-made band structures in optical lattices have inspired numerous theory proposals to use them for clean simulations of topological matter \cite{Wu:08, XJLiu:10, Gol:10, Zha:11, Sun:11, Sun:12}. In experiments with cold atoms in the lowest band of an optical lattice, topological band touching points (e.g. Dirac points) have been implemented via modulation techniques, and non-trivial Berry curvatures have been measured \cite{Jot:14, Aid:15, Duc:15}. Additional complexity is encountered in such experiments associated with engineering the necessary band structure. In higher bands topological properties are far more easily obtained, at the price that the atoms have to be excited to these bands. The lattice potential of Eq.~\ref{Eq.1} gives rise to symmetry-protected band touching points associated with the 2nd, 3rd and 4th bands, which may be viewed as topological defects in the band structure. The 4th band can be readily populated (cf. Fig.~\ref{Fig.6}) and a long-lived BEC can be formed in the $\Gamma$-point at the center of the 1st BZ, which can be used to probe the band structure properties. 

In the following, we denote the Bloch bands as $B_n$ with $n \in \{1,2,...\}$ ordered according to increasing energies. The first BZ of the lattice is replotted in Fig.~\ref{Fig.19}(a) with a path connecting the points $\Gamma$, $X_+$ and $M$. In Fig.~\ref{Fig.19}(b) the bands $B_2$, $B_3$ and $B_4$ at the $\Gamma$-point, derived from a 2D band structure calculation for a lattice with $V_{xy,0} = 7.0 \, E_{\textrm{rec}}$ and perfect $C_4$ symmetry (i.e. setting $\eta = \epsilon_{x} = \epsilon_{y} = 1$ in Eq.~\ref{Eq.1}), are plotted against $\Delta V$. At the critical value $\Delta V = \Delta V_{c}$ ($= 6.1\, E_{\textrm{rec}}$ for $V_{xy,0} = 7.0\,E_{\textrm{rec}}$) all three bands become degenerate and the $B_2$- and the $B_4$-band form a Dirac cone in the $\Gamma$-point with the locally flat $B_3$-band intersecting its origin. This can be seen in the detail on the lower edge of Fig.~\ref{Fig.19}(b) showing the involved bands within the first BZ along a trajectory connecting the points $\Gamma$, $X_{+}$, $M$, $\Gamma$ of (a). For $\Delta V >\Delta V_{c}$ ($\Delta V < \Delta V_{c}$) the $B_4$-band ($B_2$-band) separates, whereas $B_2$ and $B_3$ ($B_3$ and $B_4$) remain degenerate, thus forming a topologically protected quadratic band crossing point (TQB) \cite{Sun:09} at the centre of the first BZ. This structure is robust against changes of $\Delta V$ and $V_{xy,0}$ as long as the $C_4$ symmetry is sustained.

In experiments, discrete symmetries are usually not realized to perfection due to small symmetry-breaking perturbations. In the study of Ref.~\cite{Oel:12}, due to unavoidable imperfections of the lattice beams, the $C_4$ symmetry of the lattice is weakly broken, which acts to lift the degeneracies in Fig.~\ref{Fig.19}(b), thus leading to an avoided band crossing with an energy gap $\Delta E_{2,4}$ on the order of a small fraction of $E_{\textrm{rec}}$, as shown in Fig.~\ref{Fig.19}(c). Here, we discuss the case $\eta = 0.98$, $\epsilon_{x} = 0.93$, and $\epsilon_{y} = 0.87$. A band calculation shows that $\partial \Delta E_{2,4}/\partial V_{xy,0} < 0.02$, i.e. the gap $\Delta E_{2,4} \approx 0.26 \, E_{\textrm{rec}}$ is nearly independent of $V_{xy,0}$. On the left side of the avoided crossing ($\Delta V < \Delta V_{c}$) the bands $B_3$ and $B_4$ rapidly approach an approximately constant separation $\Delta E_{3,4} \approx 0.13 \,E_{\textrm{rec}}$, which is nearly independent of $V_{xy,0}$ ($\partial \Delta E_{3,4}/\partial V_{xy,0} < 0.002$). Within the 1st BZ the energy $\Delta E_{3,4}$ appears as the gap introduced into the TQB at the $\Gamma$-point. An analogue $V_{xy,0}$-independent gap $\Delta E_{2,3}$ arises for the bands $B_2$ and $B_3$ on the right side of the avoided crossing ($\Delta V > \Delta V_{c}$). The robustness of these gaps (with sizes on the order of ten nanokelvin) against changes of $V_{xy,0}$ and $\Delta V$ indicates that, despite the broken $C_4$-symmetry, the topological character of the bands appears to be preserved. The details on the lower edge of Fig.~\ref{Fig.19}(c) show $\Gamma X_{+} M$-trajectory plots of the bands within the first BZ away from the avoided crossing at $\Delta V = \Delta V_{c} \pm E_{\textrm{rec}}$. The insets on the upper edge show the Bloch functions at the $\Gamma$-point at the values of $\Delta V$ indicated by the blue circles. Due to the distortion of the lattice, on the left side of the avoided band crossing, the Bloch function is mainly composed of $p$-orbitals in the deep wells (indicated $\mathcal{A}$), which are slightly tilted away from the descending diagonal towards the horizontal direction. The wave function is centered around zero with nodal lines along the ascending diagonal and hence no zero momentum components arises. With increasing values of $\Delta V$ (right side of the avoided crossing) and hence increasing depths of the deep wells, the distortion becomes less pronounced and the Bloch-function develops a dominating zero-momentum component. 

\begin{figure}
\includegraphics[scale=0.4, angle=0, origin=c]{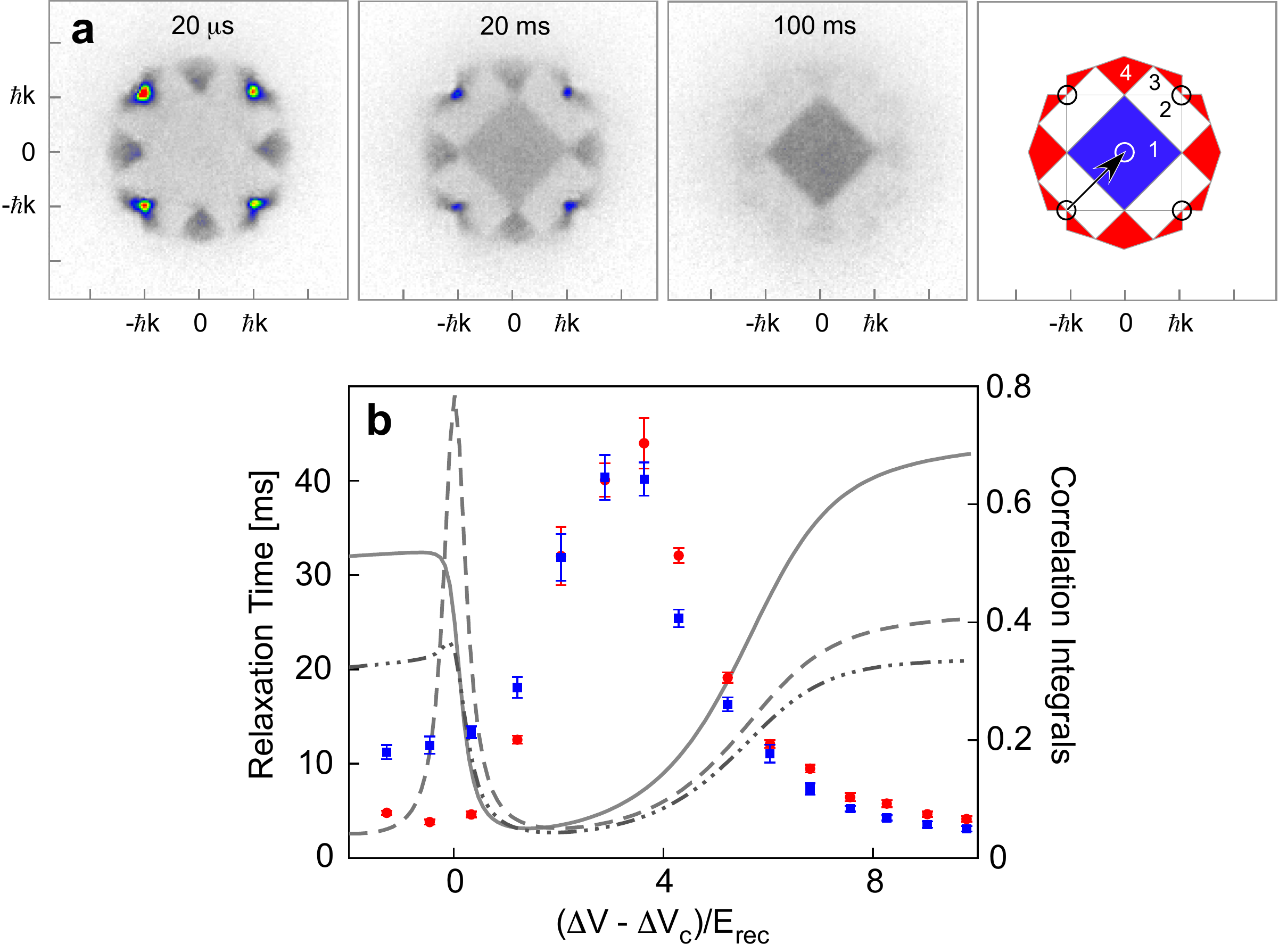}
\caption{(a) Band mapping plots recorded for $(\Delta V - \Delta V_{c})/ E_{\textrm{rec}} = 4.5$ and holding times $20\, \mu$s, $20\,$ms, $100\,$ms. In the rightmost panel a map of the theoretical BZs 1,2,3,4 is plotted. The black circles indicate the observed condensation points at the energy minima of the $B_4$-band, which connect to the centre of the 1st zone (white cricle) via reciprocal lattice vectors (black arrow). \textbf{b}, The decay time of the 4th (red disks) and the refilling time of the 1st band (blue squares) are plotted versus $\Delta V$. The grey line plots show the scaled overlap integrals $I_{1}$ (solid), $I_{2}$ (dashed), $I_{3}$ (dashed dotted) defined in Eq.~\ref{Eq.7}.}
\label{Fig.20} 
\end{figure}

The previously discussed band swapping technique permits one to prepare a BEC in the global minimum of the $B_4$-band in the $\Gamma$-point. To provide the necessary long band lifetime, $\Delta V$ is adjusted to the right hand side of the avoided crossing in Fig.~\ref{Fig.19}(b), where the $B_4$-band is well separated from all other bands (typically at $\Delta V = 9.0 \,E_{\textrm{rec}} = \Delta V_{c} + 2.9 \,E_{\textrm{rec}}$). In Fig.~\ref{Fig.20} the collision-induced relaxation of the $B_4$-band population is shown. After condensate formation, $\Delta V$ is adiabatically adjusted to some desired value within $200\,\mu$s and the atoms are held in the lattice for a variable time. Subsequent band mapping provides one with images of momentum space, in which the population of the n-th band is mapped into the n-th BZ. In Fig.~\ref{Fig.20}(a) an example is shown for $(\Delta V - \Delta V_{c})/ E_{\textrm{rec}} = 4.5$, where the effect of the avoided crossing is not relevant. Initially (for 20 $\mu$s holding time), mainly the 4th BZ is populated, with a significant fraction of the atoms residing at the equivalent quasi-momenta $(\pm \hbar k,\pm \hbar k)$, indicating condensation at the $\Gamma$-point. For larger holding times, the population of the 4th band directly decays into the 1st band (with approximately exponential time dependence), while the 2nd and 3rd BZs are not significantly involved. Only in the vicinity of the avoided crossing, a more complex decay dynamics arises with the 2nd and 3rd bands initially accumulating significant populations before finally the 1st BZ is refilled. This becomes visible in (b) (showing the relaxation times for the 4th and the 1st bands versus $\Delta V$), where close to the avoided crossing ($\Delta V \approx \Delta V_{c}$) the decay of the 4th band (red disks) is faster than the refilling of the 1st band (blue squares). The plot shows a pronounced resonance around $\Delta V \approx \Delta V_{c} + 3.5 \,E_{\textrm{rec}}$ with notably long lifetimes above 40 ms. This may be qualitatively explained by the observation that around this value of $\Delta V$ most of the atomic population resides in the local $1s$-orbits of the shallow wells, where it is protected from collisional decay, because locally no state with lower energy is available. This assertion is supported by plotting the integrals 
\begin{eqnarray} 
I_{n}(\Delta V) = \frac{\int_{\diamond}d^2r\, \rho_{4} \rho_{n}} {  \sqrt{ \int_{\diamond} d^2r \, \rho_{4}^2 \, \int_{\diamond} d^2r\,\rho_{n}^2   }    }
\label{Eq.7}
\end{eqnarray}
where $\diamond$ denotes the unit cell in configuration space, $\rho_{n}\equiv  |\phi_{n}|^2$ and $\phi_{n}, n\in \{1,2,3,4\}$ denotes the Bloch function of the n-th band for zero quasi-momentum calculated for the optical potential in Eq.~\ref{Eq.1}. These integrals measure the spatial correlations of the particle densities in the 4th and the n-th bands at the $\Gamma$-point. As shown in (b), the maximal lifetime of the 4th band arises, where these overlap integrals are small. 

\begin{figure}
\includegraphics[scale=0.45, angle=0, origin=c]{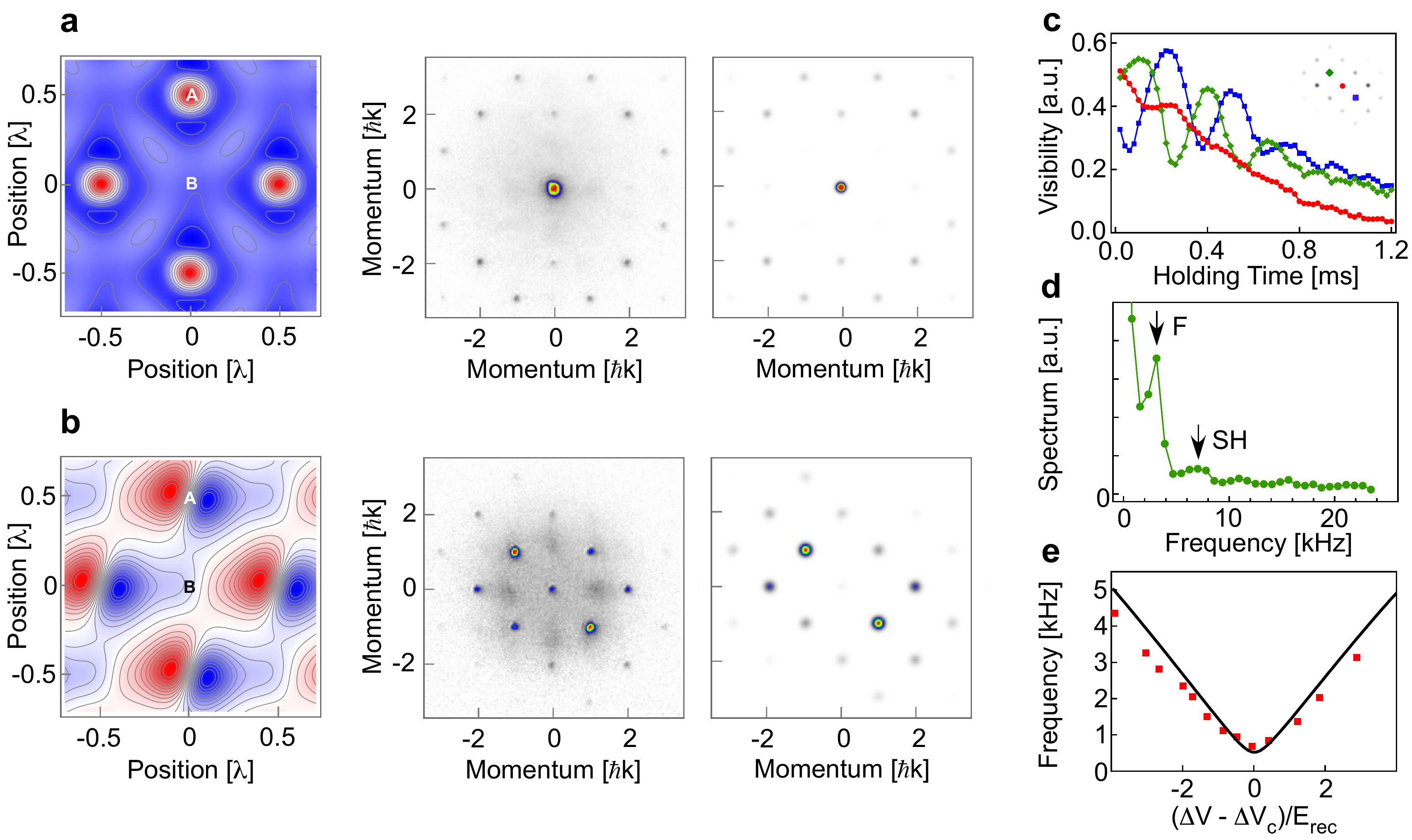}
\caption{(a) The left panel shows the Bloch function on the right side of the avoided crossing ($\Delta V = \Delta V_{c} + 9.9\,E_{\textrm{rec}}$; the same color scale as in Fig.~\ref{Fig.19}(c) applies). The center and right panels show corresponding observed and calculated momentum spectra, respectively. (b) Same as (a), however for the left side of the avoided crossing ($\Delta V = \Delta V_{c} - 1.3\,E_{\textrm{rec}}$). A sufficiently large gap ($\approx 0.8\, E_{\textrm{rec}}$) was adjusted, in order to prevent Landau-Zener dynamics as the avoided crossing is passed. (c) Landau-Zener dynamics for a gap of $\approx 0.26 E_{\textrm{rec}}$. The visibility of the $\pm (1,-1) \hbar k$-Bragg peaks (green diamonds and blue squares) and that of the $(0,0) \hbar k$-peak (red disks) are plotted versus the holding time after rapid tuning over the avoided crossing to $(\Delta V -  \Delta V_{c}) / E_{\textrm{rec}} = -3.0$. (d) Fourier spectrum of the oscillating $(-1,1) \hbar k$-peak with first and second harmonic components indicated by black arrows. (e) Red squares denote the first harmonic frequencies derived from plots as in (c). The solid line shows the calculated energy difference between the 4th and the 2nd band. For the entire graph: $V_{xy,0} = 7.8\,E_{\textrm{rec}}$, $\eta = 0.98$, $\epsilon_{x} = 0.93$, and $\epsilon_{y} = 0.87$. Parts of the figure are taken from Ref.~\cite{Oel:11}, \copyright\ American Physical Society.}
\label{Fig.21} 
\end{figure}

By tuning $\Delta V$ the condensate in the 4th band ($B_4$) may be driven across the avoided crossing at $\Delta V = \Delta V_{c}$. For small band gaps $\Delta E_{2,4}$ of less than about half a recoil energy, the rapid band relaxation occurring near the avoided crossing (cf. Fig.~\ref{Fig.20}(b)) limits one to ramping times below a millisecond, such that one operates in the non-adiabatic regime, where significant Landau-Zener dynamics is to be expected. To access the adiabatic regime, the band gap was broadened to $0.8\,E_{\textrm{rec}}$ by introducing larger deviations from $C_4$-symmetry. In this case, most atoms are maintained in the 4th band over the entire avoided crossing. As the avoided crossing is passed (with decreasing $\Delta V$, i.e, from right to left in Fig.~\ref{Fig.19}(c) the condensate wave function undergoes a dramatic change. While on the right side of the avoided crossing a momentum spectrum with approximate $C_4$-invariance with a leading zero-momentum peak is observed, on the left side the spectrum acquires finite-momentum character and reduced $C_2$-symmetry with the leading Bragg orders arising at $\pm (1,-1) \hbar k$. This is shown in Fig.~\ref{Fig.21}(a) and (b) and is discussed in the following paragraph.

In Fig.~\ref{Fig.21}(a) the case $\Delta V - \Delta V_{c} = + 9.9\,E_{\textrm{rec}}$ is shown, i.e., far to the right of the avoided crossing of Fig.~\ref{Fig.19}(c), where the $B_4$-band is well seprated from $B_2$ and $B_3$. The left hand panel in (a) shows the Bloch function of the $B_4$-band at the $\Gamma$-point, calculated from the full potential of Eq.~\ref{Eq.1} with $V_{xy,0} = 7.8\,E_{\textrm{rec}}$, $\eta = 0.98$, $\epsilon_{x} = 0.93$, and $\epsilon_{y} = 0.87$. According to the color code, which indicates nodal points by white color, this wave function is not centered around zero but possesses a large negative off-set giving rise to the extended  blue colored regions. It hence comprises a dominant zero-momentum component as is confirmed in the observed momentum spectrum plotted in the center panel in (a). Expectedly, the approximate $C_4$-symmetry of the Bloch function also shows up in this momentum spectrum. At the deep wells, marked by an $\mathcal{A}$ in (a), one finds $3s_{r^2-1}$-orbitals, which, as compared to standard $s$-orbitals, possess an extra circular nodal line, associated with large momentum components. This explains why the higher order Bragg peaks, as for example $(2,2)\, \hbar k$, dominate in the momentum spectrum. The right hand panel in (a) shows a momentum spectrum calculated according to the Bloch function in the left hand panel. The agreement with the observation in the central panel appears excellent. The physics on the left side of the avoided crossing ($\Delta V - \Delta V_{c} = - 1.3\,E_{\textrm{rec}}$) is discussed in Fig.~\ref{Fig.21}(b). Here, the bands $B_3$ and $B_4$ run parallel to each other  with a tiny energy distance $\Delta E_{3,4}$ due to the lattice distortion present in the experiment. Perfect $C_4$-symmetry would imply $\Delta E_{3,4}=0$ and hence complete degeneracy of $B_3$ and $B_4$. In this case the deep wells would host degenerate $p_x$- and $p_y$-orbitals. For non-zero $\Delta E_{3,4}$ according to the parameter choice in Fig.~\ref{Fig.21}, the degeneracy is lifted in favor of $p$-orbitals tilted with respect to the $x$-axis by an angle $-14^{\circ}$. This is seen in the Bloch function of the 4th band at the $\Gamma$-point, shown in the left hand panel in Fig.~\ref{Fig.21}(b). The Bloch function exhibits a pure standing wave geometry with an orientation along the descending diagonal. This explains the observed $C_2$-symmetry and the vanishing zero momentum component in the corresponding observed and calculated momentum spectra displayed in the central and right hand panels of (b), respectively. 

\begin{figure}
\includegraphics[scale=0.5, angle=0, origin=c]{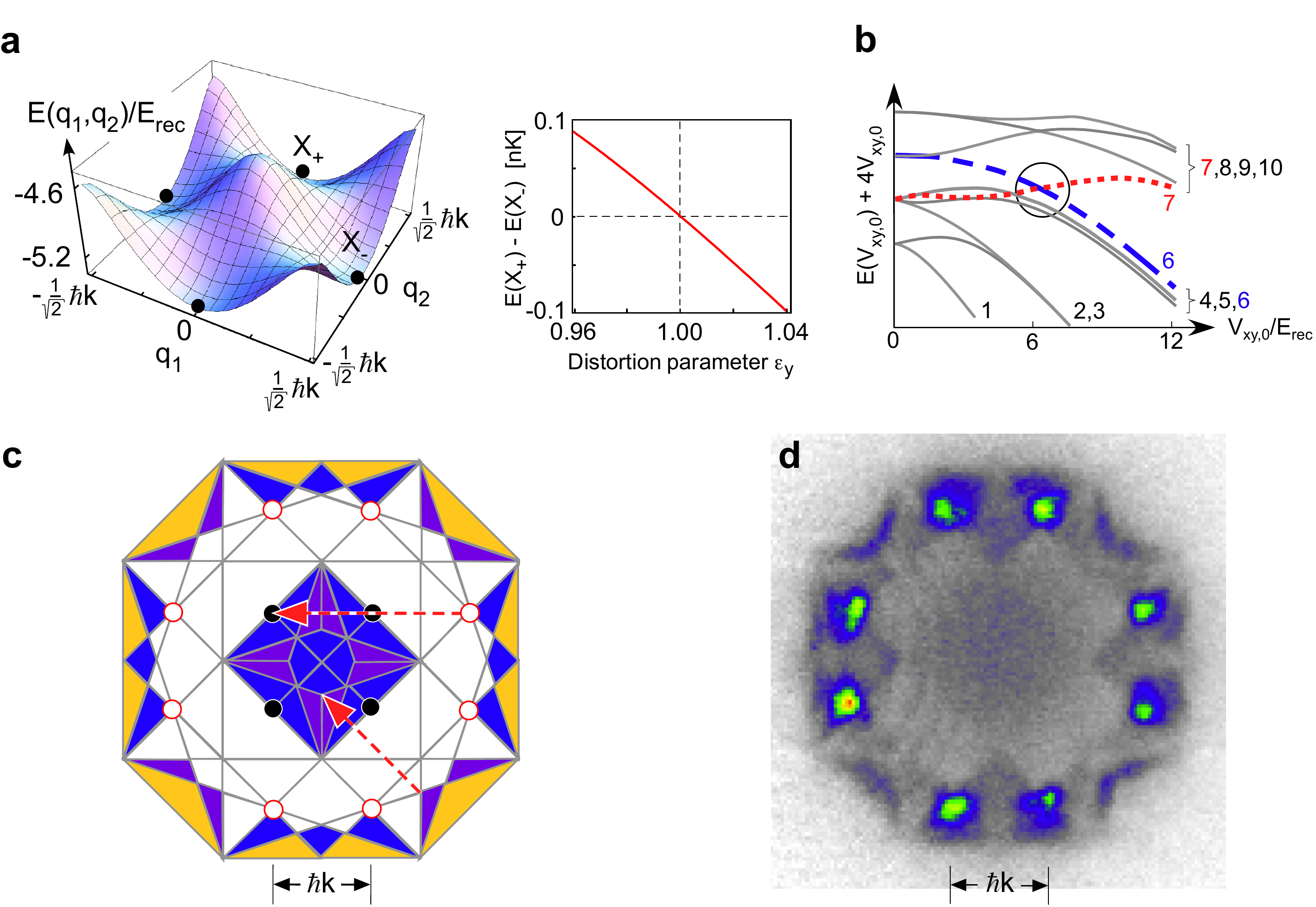}
\caption{(a) Left panel: The 7th band possesses two nearly degenerate minima at $X_{\pm}$ (black disks). Right panel: The energy difference $E(X_{+}) - E(X_{-})$ between the condensation points $X_{+}$ and $X_{-}$, calculated from the potential in Eq.~\ref{Eq.1} with $V_{xy,0} = 8.3\,E_{\textrm{rec}}$ and $\theta = 0.66\,\pi$, is plotted in units of nanokelvin versus the distortion parameter $\epsilon_{y}$. (b) The energies at the $X_{\pm}$-points of the bands 1-10 are plotted versus $V_{xy,0}$ (with $\epsilon_{y}=1$ and $\theta = 0.66\,\pi$). Note the band crossing between the 6th and 7th bands slightly above $V_{xy,0} = 6\,E_{\textrm{rec}}$. (c) Schematic of the 6th (dark blue and purple areas) and 7th (light orange areas) BZs. The eight red circles mark the energy minima within the 6th BZ. In the center the first BZ is reconstructed by translations of subsets of the 6th BZ via reciprocal lattice vectors indicated by the red dashed arrows. (d) Observed populations of different Brillouin zones after applying the population swapping procedure with final values of $\theta = 0.66\, \pi$ and $V_{xy,0} = 8.3\,E_{\textrm{rec}}$. Parts of the figure are taken from Ref.~\cite{Oel:11}, \copyright\ American Physical Society.}
\label{Fig.22} 
\end{figure}

The non-adiabatic case was studied for $\Delta E_{2,4} \approx 0.26 \, E_{\textrm{rec}}$ corresponding to 25~nanokelvin. After the preparation of the condensate at $\Delta V = \Delta V_{c} + 2.9 \,E_{\textrm{rec}}$, values $\Delta V < \Delta V_{c}$ on the left side of the avoided crossing are adjusted within $400\,\mu$s, and the time evolution of the momentum spectrum is recorded. An example for $(\Delta V -  \Delta V_{c}) / E_{\textrm{rec}} = -3.0$ is shown in Fig.~\ref{Fig.21}(c), where the visibility of the $\pm (1,-1) \hbar k$-Bragg peaks (green diamonds and blue squares) and that of the $(0,0) \hbar k$-peak (red disks) are plotted versus the holding time after the jump over the avoided crossing. Significant oscillations are observed, which can be qualitatively modeled as a beat between the single-particle Bloch functions of the 4th and 2nd band at the $\Gamma$-point. A closer inspection shows that the frequency spectra of these oscillations comprise small second-harmonic components as is illustrated for the $(-1,1) \hbar k$-peak in Fig.~\ref{Fig.21}(d), which is expected as a result of the non-linearity introduced by collisional interactions. The second-harmonic contributions give rise to slightly increased (decreased) curvatures in the minima (maxima) of the oscillations in Fig.~\ref{Fig.21}(c). Collisional relaxation is also responsible for the observed decay of the visibility. If an additional lattice potential along the $z$-direction is applied in order to increase the collision energy per particle, one finds correspondingly decreased decay times. Evaluation of curves similar as in Fig.~\ref{Fig.21}(c) for variable $\Delta V$ yields the red squares in Fig.~\ref{Fig.21}(e). The solid trace repeats the energy difference between the 4th and the 2nd band from Fig.~\ref{Fig.19}(b) obtained from a numerical band structure calculation using the potential of Eq.~\ref{Eq.1} and the carefully measured values of $\eta$, $\epsilon_{x}$ and $\epsilon_{y}$. Despite neglecting collisions, the calculations without the use of fitted parameters well approximate the observations. The small deviations of the observed frequencies towards values slightly below the single-particle calculations cannot be reduced by choosing different values for $\eta, \epsilon_{x}, \epsilon_{y}$, but rather indicate the effect of collisions. 

\section{Chiral superfluid order in the seventh band}
Beyond the 4th band, the next higher band in a 2D lattice geometry that can be selectively populated with the population swapping technique is the 7th band. This has been experimentally explored in Ref.~\cite{Oel:11}. The 7th band, similarly as the 2nd band, possesses two inequivalent minima at the $X_{\pm}$-points (cf. left panel of Fig.~\ref{Fig.22}(a)). A numerical band calculation shows that in comparison to the 2nd band the degeneracy between the two band minima is significantly more robust with respect to lattice distortions. The tuning of the energy difference between $X_{+}$ and $X_{-}$ with the distortion parameter $\epsilon_{y}$ is illustrated in the right panel of Fig.~\ref{Fig.22}(a). Comparison to the situation in the 2nd band in Fig.~\ref{Fig.9}(b) shows a nearly ten-fold decreased sensitivity. This is confirmed by the observation that in the entire range of experimentally accessible lattice distortions both condensation points are nearly equally populated. The 7th and the 6th bands cross, as $V_{xy,0}$ is adiabatically tuned to zero in the context of the band mapping technique, as is shown in Fig.~\ref{Fig.22}(b), such that the population of the 7th band is largely transferred to the 6th BZ. The 6th and 7th BZs are illustrated in Fig.~\ref{Fig.22}(c). The energy minima within the 6th BZ are highlighted by red circles. These points are equivalent to the $X_{\pm}$-points in Fig.~\ref{Fig.22}(a) modulo reciprocal lattice vectors indicated in (c) by red dashed arrows. This is where the atoms should gather if BEC occurs, which is in accordance with the observed band mapping image in (d). This image was taken immediately after populating the 7th band. Nevertheless, accumulation of atoms in the $X_{\pm}$-points is already visible due to the fast tunneling rates close to the continuum.

\begin{figure}
\includegraphics[scale=0.5, angle=0, origin=c]{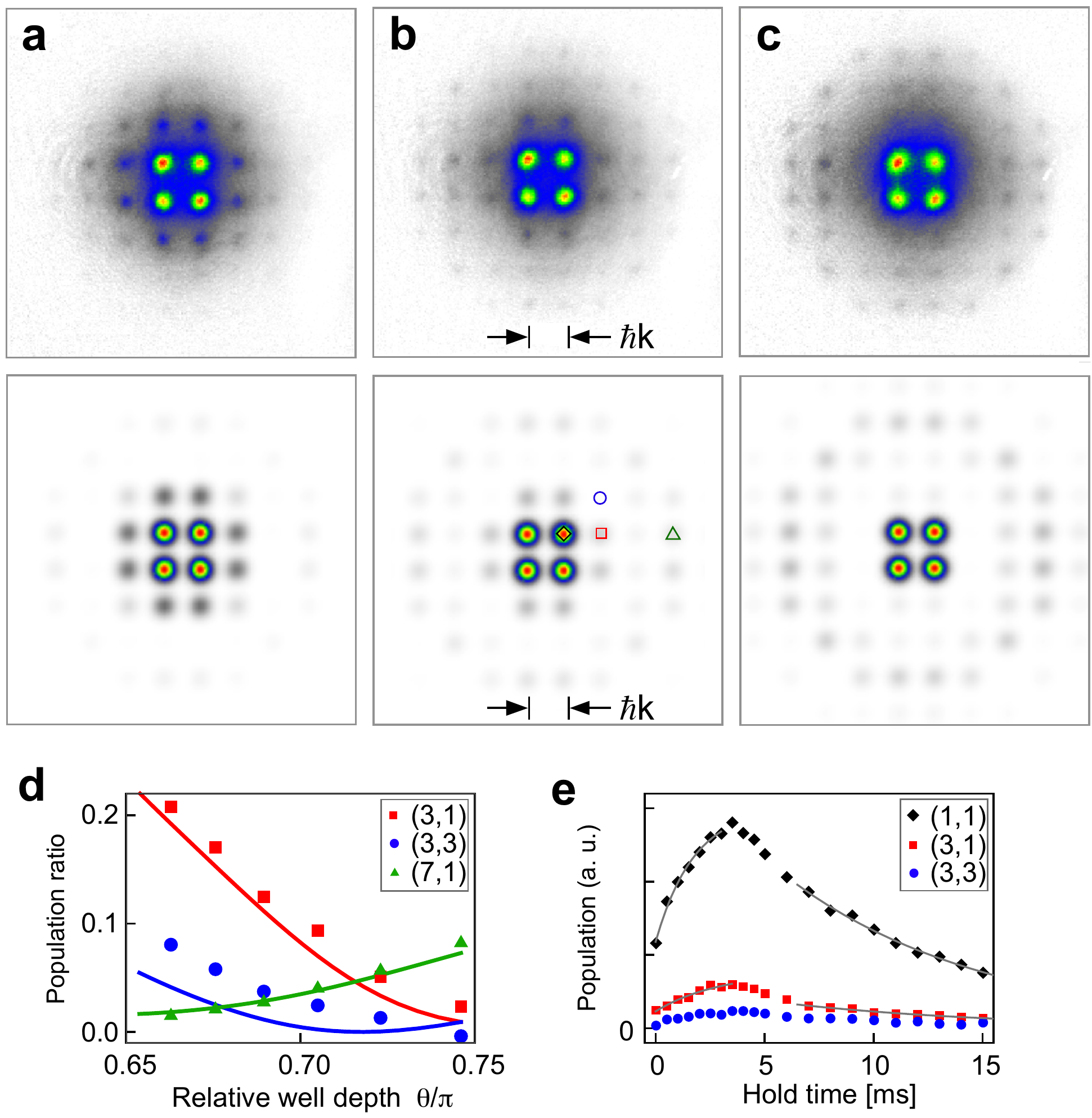}
\caption{In (a), (b), and (c) momentum spectra are shown for a hold time of 1 ms, $V_{xy,0} = 8.3\,E_{\textrm{rec}}$ and $\theta/\pi = 0.66, 0.70, 0.75$, respectively. Observations (calculations) are shown in the upper (lower) row. (d) Populations of the Bragg peaks (3,1), (3,3), and (7,1) normalized to that of the (1,1) peak plotted versus $\theta$. The symbols show observations, the solid lines show the corresponding results of a band calculation. The symbols are repeated in the calculated spectrum in (b) for identification of the relevant Bragg peaks. (e) Temporal evolution of the populations of Bragg peaks (1,1), (3,1), and (3,3). The solid lines are exponential fits applied in the wings of the graphs to determine relaxation times. Reproduced from Ref.~\cite{Oel:11}, \copyright\ American Physical Society.}
\label{Fig.23} 
\end{figure}

Momentum spectra are shown in Fig.~\ref{Fig.23}(a), (b), and (c) with observations taken after a hold time of 1~ms in the upper row and corresponding numerical calculations in the lower row. Even details in the distribution of populations of higher Bragg resonances are well reproduced by the calculations. This is seen in a more quantitative presentation in Fig.~\ref{Fig.23}(d), which plots the populations of the (3,1)-, (3,3)-, and (7,1)-Bragg peaks normalized to that of the (1,1)-peak versus $\theta$ in Eq.~\ref{Eq.1}. The observations are indicated by the symbols and the calculations are given by the solid lines. In Fig.~\ref{Fig.23}(e) the emergence of the condensate and its depletion due to band relaxation is illustrated by plotting the temporal evolution of the populations of Bragg peaks (1,1), (3,1), and (3,3). The solid lines are exponential fits applied in the wings of the graphs to determine the relaxation times (1.8 ms and 9.1 ms for the (1,1)-peak (black diamonds) and 2.9~ms and 10.1~ms for the (3,1)-peak identified by red rectangles).

\begin{figure}
\includegraphics[scale=0.28, angle=0, origin=c]{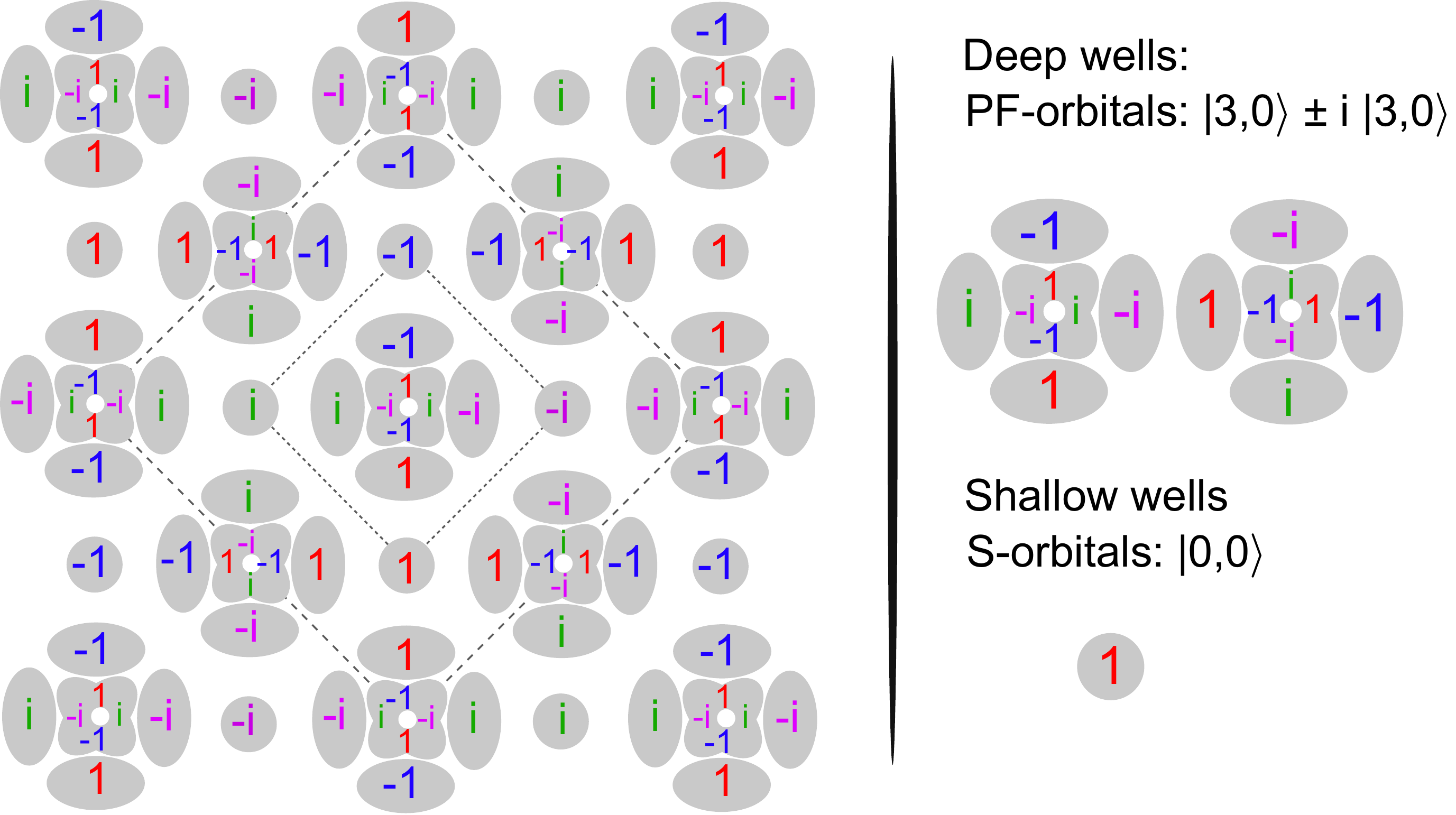}
\caption{\label{Fig.24} Orbital configuration of the order parameter. The grey areas characterize the antinodal structure of the orbitals, the colored numbers indicate the local phases. Reproduced from Ref.~\cite{Oel:11}, \copyright\ American Physical Society.}
\end{figure}

As confirmed by a band calculation, the ground state wave function in the 7th band is composed of local $s$-orbitals in the shallow wells and local $pf_{2x^3-3x^2}$ and $pf_{2y^3-3y^2}$ hybrid orbitals in the deep wells, which in harmonic approximation display a spatial $\frac{1}{\sqrt{3}}\,(2x^3-3x^2)\,e^{-(r/2\sigma)^2}$- and $\frac{1}{\sqrt{3}}\,(2y^3-3y^2)\,e^{-(r/2\sigma)^2}$-dependence with $r^2 \equiv x^2+y^2$ and $\sigma$ denoting the radius of the associated $s$-orbitals. These hybrid orbitals, which coincide with the maximally stretched cartesian orbitals of the 2D harmonic oscillator $|3,0\rangle$ and $|0,3\rangle$ with three oscillation quanta, are energetically most favorable due to the negative anharmonicity of the sinusoidal lattice potential. Their relative phase is fixed via repulsive collisions to $\pi /2$ in order to maximize the mode volume so that the atoms can best avoid each other. Kinetic energy favors equal local phases across the tunneling junctions, giving rise to the wave function schematically illustrated in Fig.~\ref{Fig.24}, which as its analog in the 2nd band in Fig.~\ref{Fig.14}(b) displays orbital and plaquette currents. As a consequence the translation symmetry of the lattice and time-reversal symmetry are broken. It is interesting to note that, in contrast to the case of the second band, the local orbitals $pf_{2x^3-3x^2} \pm i \, pf_{2y^3-3y^2}$ formed in the deep wells do not possess sharp angular momentum but are composed of $p$- and $f$-orbitals according to $pf_{2x^3-3x^2} =  \frac{\sqrt{3}}{2} p_{x(r^2-2)} + \frac{1}{2} f_{x(x^2-3y^2)}$ and $pf_{2y^3-3y^2} = \frac{\sqrt{3}}{2} p_{y(r^2-2)} + \frac{1}{2} f_{y(y^2-3x^2)}$. In particular, angular momentum is not maximized and hence Hund's second rule does not apply here \cite{Wu:09}. 

\section{Conclusion}
Orbital degrees of freedom are ubiquitous in electronic condensed matter, giving rise to the rich phenomenology of transition metal oxides and other highly complex materials with unconventional intrinsic symmetries and topologies. In condensed matter, orbital degrees of freedom usually come together with other degrees of freedom associated with spin and charge, such that an exceedingly complex physical scenario arises, which typically is difficult to analyze. Optical lattices provide an ideal experimental arena to selectively prepare and study orbital degrees of freedom in a tailored environment, where competing mechanisms also yielding emergent complexity can be kept under control. This may help to promote our understanding of electronic complex matter, however, in particular, if bosonic atoms are used, optical lattices also allow one to engineer novel many-body scenarios beyond what is possible in condensed matter, which are highly interesting in their own right. In this review, we have summarized our work during the past few years on the preparation of bosonic atom samples condensed in higher bands of optical lattices. We have shown that bipartite square lattice geometries allow one to realize the necessary requirements: control of band relaxation, selective excitation of the atoms to the target band, efficient cross-dimensional tunneling in the target band. We have explored various physical scenarios in the 2nd, the 4th and the 7th band, however, different choices of the target bands or of the lattice geometry should be possible. Only a few examples have been considered yet, which however clearly illustrate the wealth of new physics arising when orbital degrees of freedom come into play in optical lattice systems. Orbital degrees of freedom provide two most interesting consequences: directionality of tunneling and interactions, and orbital degeneracy. As a consequence of the latter, even small interactions can determine the ground state of the system. In our work with bosonic atoms in the second band of a bipartite square lattice, orbital degeneracy is the basis for the emergence of interaction-induced chiral superfluidity. An obvious next step is to also study orbital physics in optical lattices with fermions. Excited bands can come with interesting topological structure, which in addition to the physics associated with broken symmetries can give rise to exotic phenomena like, for example, topologically protected chiral edge states. Orbital optical lattices open up new avenues to experimentally study such phenomena also with bosons, which is not possible in electronic condensed matter.

\section{acknowledgments}
This work was partially supported by DFG-SFB 925 and the Hamburg Centre for Ultrafast Imaging (CUI). AH is grateful to W. Vincent Liu, Cristiane Morais Smith, Arun Paramekanti, Ludwig Mathey, and Claus Zimmermann for discussions and to the entire team of the Wilczek quantum center for the outstanding hospitality.

\end{document}